\documentclass[letterpaper,oneside]{article}
\usepackage{graphicx}
\usepackage{amsmath,amssymb,amsbsy,float}
\begin{document}

\begin{titlepage}
\title{Electrical and magnetic properties of nano-scale  $\pi$-junctions}
\author{Samanta Piano}
\date{\it Physics Department, CNR-Supermat Laboratory, University of Salerno, Via S. Allende, 84081 Baronissi (SA), Italy}

\end{titlepage}

\maketitle

\bigskip

\begin{abstract}

The physics of the $\pi$ phase shift in ferromagnetic Josephson
junctions enables a range of applications for spin-electronic
devices and quantum computing. In this respect our research is
devoted to the evaluation of the best materials for the development
and the realization of the quantum devices based on superconductors
and at the same point towards the reduction of  the size of the
employed heterostructures towards and below nano-scale. In this
chapter we report our investigation of transitions from "0" to
"$\pi$" states in  Nb Josephson junctions with strongly
ferromagnetic barriers of Co, Ni, Ni$_{80}$Fe$_{20}$ (Py) and Fe. We
show that it is possible to fabricate nanostructured Nb/ Ni(Co, Py,
Fe)/Nb $\pi$-junctions with a nano-scale magnetic dead layer and
with a high level of control over the ferromagnetic barrier
thickness variation. In agreement with the theoretical model we
estimate, from the oscillations of the critical current as function
of the ferromagnetic barrier thickness, the exchange energy of the
ferromagnetic material and we obtain that it is close to bulk
ferromagnetic materials implying that the ferromagnet is clean and
S/F roughness is minimal. We conclude that S/F/S Josephson junctions
are viable structures in the development of superconductor-based
quantum electronic devices;
 in particular Nb/Co/Nb and Nb/Fe/Nb multilayers with their low value of the magnetic dead layer and high value of the exchange energy can
 readily be used in controllable two-level quantum information systems. In this respect, we discuss
applications of our nano-junctions to engineering magnetoresistive devices such as programmable pseudo-spin-valve
Josephson structures.

\end{abstract}

\clearpage

\section{Introduction}

Conventional electronics is based on the transport of electrical
charge carriers, however the necessity to have more versatile and
efficient devices on micro- and nano-scale has diverted the
attention towards the spin of the electron rather than its charge:
this is the essence  of {\em spintronics}.   Although spintronic
devices were traditionally built with  semiconductors and/or
ferromagnetic materials, nowadays hybrid structures constituted by
superconductors (S) and ferromagnetic (F) materials are emerging as
promising alternatives. They serve as candidates  to realize and
investigate systems useful both as testgrounds to probe fundamental
physics questions, and as realistic elements for application in the
spintronics industry thanks to the improved degree of control on the
spin polarized current.

 In hybrid S/F structures it has been verified that,
 due to the simultaneous presence of these two competitive orders,
peculiar effects appear: the superconductivity is reduced by the
spin polarization of the F layer, while due to the proximity effect
the Cooper pairs can enter in the superconductor resulting in a
oscillatory behavior of the density of states. The last effect opens
the doors towards new and exciting developments for the realization
of quantum electronic devices. In fact in S/F/S systems, due to the
oscillations of the order parameter, some oscillations manifest in
the Josephson critical current as a function of the ferromagnetic
barrier thickness evidencing the presence of two different states, 0
and $\pi$, corresponding to the sign change of the Josephson
critical current.

Our research, presented in this chapter, fits in this rapidly
developing and exciting area of condensed matter physics  \cite{Samthesis}. Within the
context of the spin polarized devices based on the interplay between
superconductivity and ferromagnetism, we have fabricated S/F/S
nano-structured Josephson junctions constituted by a low temperature
superconductor, Niobium (Nb), and strong ferromagnetic metal,
Nickel(Ni), Ni$_{80}$Fe$_{20}$ (Py), Cobalt (Co) and Iron (Fe), and
we have investigated their magnetic and electrical properties. These
structures have evidenced a small magnetic dead layer and
oscillations of the critical current as a function of the
ferromagnetic barrier. These oscillations show excellent fits to
existing theoretical models. We also determine the Curie temperature
for Ni, Py and Co. In the case of Co and Fe we estimate the mean
free path to confirm that the oscillations are in the clean limit
and from the temperature dependence of the $I_cR_N$ product we show
that its decay rate exhibits a nonmonotonic oscillatory behavior
with the ferromagnetic barrier thickness.  We investigate the presence of the Shapiro steps on the $I$ \emph{vs} $V$ curve applying a microwave and the effect of the magnetic field on the maximum supercurrent. In this last case we find oscillations of the maximum of the supercurrent corrisponding to a Fraunhofer pattern. Finally, in
the case of Co barrier, focusing in detail on a single 0-$\pi$ phase
transition we show evidence for the appearance of a second harmonic
in the current-phase relation at the minimum of the critical
current. For the high value of the exchange energies and small
magnetic dead layer the S/F/S structures with Co or Fe barrier can
be considered as good candidates for the realization of quantum
devices.

\section{F/S Junctions: Basic Aspects}\label{FSsection}

The aim of this section is to give an outline of the physics of the
Ferromagnetic-Superconducting (F/S) interfaces (for a review see
\cite{Buzdin}).
\begin{figure}[t!]
\centering \includegraphics[width=10cm]{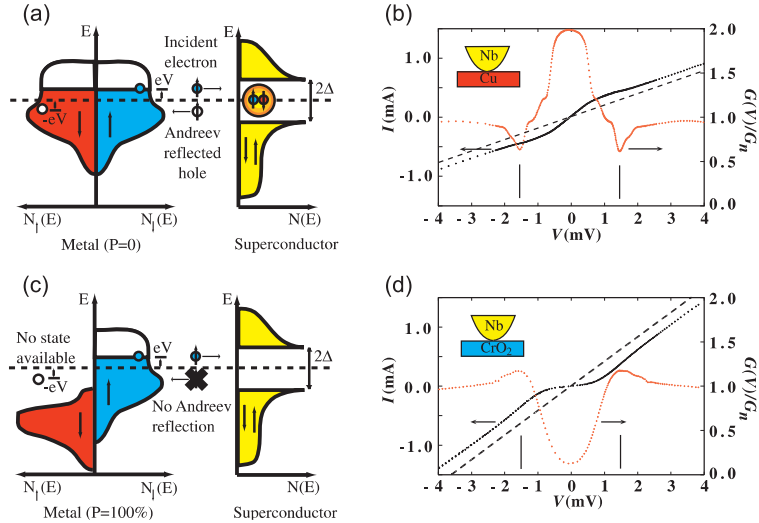}
\caption{\label{soulen} Andreev reflection process for a metal with
spin polarization $P=0$ (a) and $P=100\%$ (c); experimental measurements of a N/S
interface (b) and F/S junction (d). Figure adapted from
Ref.\cite{Soulen}.}
\end{figure}
Andreev reflection plays an important role to
understand the transport process in F/S junctions. The Andreev
reflection near the Fermi level preserves energy and momentum but
does not preserve spin, in other words the incoming electron and the
reflected hole have opposite spin. This is irrelevant for the
transport in N/S junctions (N stands for normal metal) due to the spin-rotation symmetry. On the
other hand at F/S interfaces, since the spin-up and spin-down bands
in F are different, the spin flipping changes the conductance
profile. In particular in fully spin-polarized metals all carriers
have the same spin and the Andreev reflection is totaly suppressed;
therefore, at zero bias voltage the normalized conductance becomes
zero (see Fig. \ref{soulen}). In general, for arbitrary polarization
$P$, it can easily be shown that
\begin{equation}\label{polarization}
\frac{G_{FS}(0)}{G_{FN}}=2(1-P)
\end{equation}
where, in terms of the spin-up $N_\uparrow$ and spin-down
$N_\downarrow$ electrons, the spin polarization is defined as
\begin{equation}
P=\frac{(N_\uparrow -N_\downarrow)}{(N_\uparrow + N_\downarrow)},
\end{equation}

From this analysis it has been shown that the point contact
measurements based on the Andreev reflection process give a
quantitative estimation of the polarization of the ferromagnetic
material \cite{Soulen,Strijkers}. In fact from the reduction of the
zero bias conductance peak it is possible with a modified
Blonder-Tinkham-Klapwijk model
to estimate the polarization $P$.

\begin{figure}[t!]
\centering \includegraphics[width=5cm]{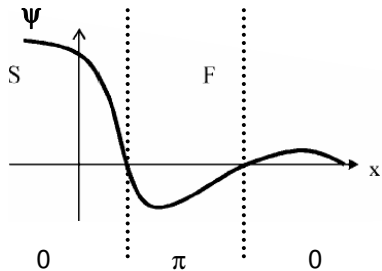}
\caption{\label{oscillpsiFS} Oscillatory behavior of the
exponential decay of the superconducting order parameter at the
F/S interface.}
\end{figure}

In a ferromagnet in proximity to a superconductor, the coherence
length $\xi_F$, due to the presence of the exchange field $E_{ex}$,
is given by:
\begin{equation}
\xi_F=\sqrt{\frac{D\hbar}{2(\pi K_BT+iE_{ex})}},
\end{equation}
where $D$ is the diffusive coefficient. If the exchange energy is
large compared to the temperature, $E_{ex} > k_BT$, the coherence
length is much shorter than in the case of the N/S proximity effect.
In addition to the reduced coherence length compared to typical N/S
structures, a second characteristic property arises from the complex
nature of the coherence length: the induced pair amplitude
oscillates spatially in the ferromagnetic metal as a consequence of
the exchange field acting upon the spins of the two electrons
forming a Cooper-pair \cite{B2,B3,R1,R2} (see Fig.
\ref{oscillpsiFS}). This oscillation includes a change of sign and
by using appropriate values for the exchange energy and F-layer
thickness, negative coupling can be realized. We
refer to the state corresponding to a positive sign of the real part
of the order parameter as ``$0$-state'' and that corresponding to a
negative sign of the order parameter as ``$\pi$-state''.

Demler et al. \cite{Demler} gave a qualitative picture of this
oscillatory character. They considered a Cooper pair transported
adiabatically across an F/S interface, with its electron momenta
aligned with the interface normal direction.
 The pair entering in the F region decays exponentially on the length scale of the normal metal coherence length.
 Then the up-spin electron, oriented along
the exchange field, decreases its energy by $h=E_{ex}/\hbar$, where
$E_{ex}$ is the exchange energy of the F layer. On the other hand,
the down-spin electron increases its energy by $E_{ex}$. To
compensate this energy variation, the up-spin electron increases its
kinetic energy, while the down-spin electron decreases its kinetic
energy. As a result the Cooper pair acquires a center-of-mass
momentum $Q=2E_{ex}/v_F$, which implies the modulation of the order
parameter with period $\pi v_F/E_{ex}$, where $v_F$ is the Fermi
velocity.
\begin{figure}[t!]
\centering \includegraphics[width=6cm]{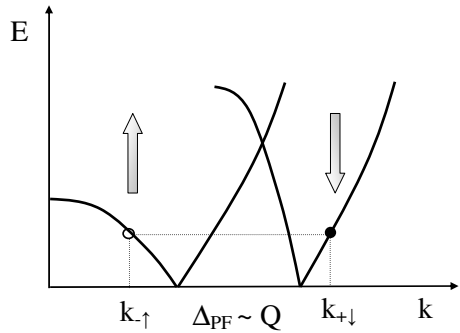} \caption{\label{FS}
Schematic of the Andreev reflection process in a F/S junction, the
momentum shift, $\Delta_{pF}$, is dominated by the spin splitting of
the up and down bands.}
\end{figure}

As a consequence of the oscillations of the order parameter, a
similar oscillatory behavior is observed for the density of states.
This behavior can be explained by considering the spin effect on the
mechanism of Andreev reflections \cite{Buzdin}.

The process is illustrated in Fig. \ref{FS} using the
energy-momentum dispersion law of the normal metal: in the case of a
N/S interface an incoming electron in a normal metal N with energy
lower than the superconducting energy gap $\Delta$ from the Fermi
level can be reflected into a hole at the N/S interface (Andreev
Reflection) \cite{Andreev}. If the normal layer is very thin the
density of states in N is close to that of the Cooper pair
reservoir. The situation is strongly modified if the normal metal is
ferromagnetic. As Andreev reflections invert spin-up into spin-down
quasiparticles and vice versa, the total momentum difference
includes the spin splitting of the conduction band: $\triangle
P_F\simeq Q$. The density of states is modified in a thin layer on
the order of $\xi_F$. In particular, the interference between the
electron and hole wave functions produces an oscillating term in the
superconducting density of states with period $E_{ex}/\hbar v_F$.
This effect has been observed experimentally by Kontos et al.
\cite{Kontos}

\subsection{Theory of the Josephson $\pi$-junctions}

Due to the spatially oscillating induced pair amplitude in F/S
proximity structures it is possible to realize negative coupling of
two superconductors across a ferromagnetic weak link (S/F/S
Josephson junctions). In this case of negative coupling, the
critical current across the junction is reversed when compared to
the normal case giving rise to an inverted current-phase relation.
Because they are characterized by an intrinsic phase shift of $\pi$,
these junctions are called {\em Josephson $\pi$-junctions}.

One of the manifestations of the $\pi$ phase is a non-monotonic
variation of the critical temperature and the critical current, with
the variation of the ferromagnetic layer thickness ($t_F$)
\cite{Buzdin}. Referring to the current-phase relation $I_c=I_0 \sin
\phi$, for a S/I/S Josephson junction the constant $I_0>0$ and the
minimum energy is obtained for $\phi=0$. In the case of a S/F/S
Josephson junctions the constant $I_0$ can change its sign from
positive to negative indicating the transition from the $0$-state to
$\pi$-state. Physically the changing in sign of $I_0$ is a
consequence of a phase change in the electron pair wave function
induced in the F layer by the proximity effect. Experimentally,
measurements of $I_c$ are insensitive to the sign of $I_0$ hence the
absolute value of $I_0$ is measured, so we can reveal a non-monotonic
behavior of the critical current as a function of the F layer
thickness. The vanishing of critical current marks the transition
from $0$ to $\pi$ state. The dependence of the critical current on
the thickness of the ferromagnetic layer in S/I/F/S junctions has
been experimentally investigated by Kontos et al. \cite{Kontos2002}. The
quantitative analysis of the S/F/S junctions is rather complicated,
because the ferromagnetic layer can modify the superconductivity at
the F/S interface. Then other parameters, such as the boundary
transparency, the electron mean free path, the magnetic scattering,
etc can affect the critical current.  It is outside the purpose of
this chapter to derive explicitly the expression of the
critical current as a function of the F layer, for the interested
reader we refer to this review \cite{Buzdin}.

 The majority of experimental studies have concentrated
on weak ferromagnets where $E_{ex}\sim K_B T_c$, where $T_c$ is the
superconducting critical temperature, resulting in multiple
oscillations in $I_c$ with temperature and $t_F$. In the case of
strong ferromagnets, where $T_{\rm Curie} \gg T_c$, only
oscillations of $I_c$ with $t_F$, and not with temperature, are
observed. In this chapter we will present the study of the
oscillations of the critical current as a function of $t_F$ for
S/F/S Josephson junctions with {\em strong} ferromagnetic barriers.

The generic expression of the critical current as a function of F
layer is given by:
\begin{equation}
\ I_c R_N (t_F)= I_c R_N (t_0)\Bigg | \frac{\sin
\frac{t_F-d_1}{\xi_{2}}}{\sin \frac{t_F-t_0}{\xi_{1}}} \Bigg| \exp
\bigg\{  \frac{t_0-t_F}{\xi_{1}} \bigg\}, \label{general}
\end{equation}
where $t_1$ is the thickness of the ferromagnet corresponding to
the first minimum and $I_c R_N (t_0)$ is the first experimental
value of $I_c R_N$ ($R_N$ is the normal state resistance), and
$\xi_1$ and $\xi_2$ are the two fitting parameters.  Eq.
\ref{general} ranges in the clean and in the dirty limit. In
particular, in clean limit, $t_F<L$ where L is the mean free path of
the F layer, $\xi_2= v_F \hbar/2 E_{ex}$. In this way, known $\xi_2$
and estimating the Fermi velocity from reported values in
literature, one can calculate the exchange energy of the
ferromagnetic barrier.

Then, in the case of clean limit the oscillations of $I_CR_N$
\emph{vs} $t_F$ can be modeled by a simpler theoretical model
\cite{Buzdin82} given by:
\begin{equation}
\ I_c R_N(t_F) = I_c R_N (t_0) \frac{\mid \sin(2 E_{ex} t_F /\hbar
v_f) \mid}{2 E_{ex} t_F /\hbar v_f}, \label{cleanlimit}
\end{equation}
where in this case the two fitting parameters are $v_F$ and
$E_{ex}$.

 On the other hand, in dirty limit $t_F>L$ we can model the
oscillations \cite{Bergeret} by the following formula:
\begin{equation}
\ I_c R_N(t_F) = I_c R_N (t_0) \mid Re \sum_{\omega_m >0}
\frac{\Delta^2}{\Delta^2 + \omega_m^2} \int^1_{-1}
\frac{\mu}{\sinh(k_\omega t_F / \mu L)}d\mu \mid, \label{dirtylimit}
\end{equation}
where $\Delta$ is the superconducting order parameter, $\omega_m$ is
the Matsubara frequency and is given by $\omega_m = \pi T k_B
(2m+1)$ where $T$ is the transmission coefficient and $m$ is an
integer number. $k_\omega=(1+2\mid \omega_m \mid \tau / \hbar) -
2iE_{ex} \tau/\hbar$ and $\mu=\cos \theta$ where $\theta$ is the
angle the momentum vector makes relative to the distance normal to
the F/S interface. $L$ is given by $v_F \tau$ and $\tau$ is the
momentum relaxation time. In this case the fitting parameters are
$v_F$, $E_{ex}$ and the mean free path $L$ of the ferromagnetic layer.

\section{Josephson Junction Fabrication}\label{SFShetero}

\begin{figure}[!ht]
\centering \includegraphics[width=5cm]{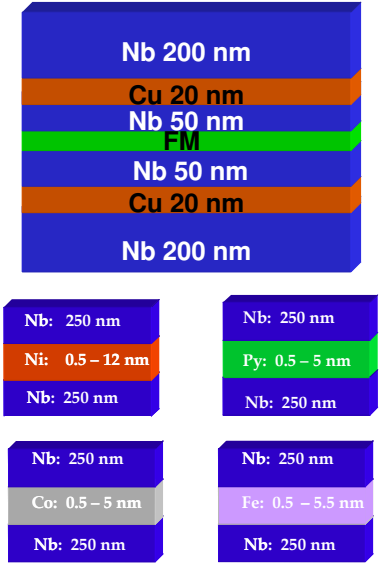}\\
   \caption{A schematic picture of the S/F/S heterostructure used in this chapter.}\label{heterostructures}
\end{figure}

To investigate the physics of the ferromagnetic $\pi$-junctions and
their possible applications for spin-electronic devices we have
realized and characterized Nb (250 nm thick) / Ferromagnetic layer /
Nb (250 nm thick) Josephson junctions. As ferromagnetic layer we
have used: Ni, Py, Co and Fe. In the case of Co, Fe and Py their
respective thicknesses $t_{\rm F}$ were varied from
$\simeq0.5$-$5.5$ nm while, in the case of Ni, $t_{\rm Ni}$ was
varied from $\simeq1.0$ to $10$ nm. To assist processing in a
focused ion beam (FIB) microscope, a 20 nm normal metal interlayer
of Copper (Cu) was deposited inside the outer Nb electrodes, but 50
nm away from the Fe barrier. We remark that the 20 nm of Cu is a
thickness smaller than its coherence length, so it is completely
proximitized into the Nb, and it does not affect the transport
properties of the Josephson junction. Refer to Fig.
\ref{heterostructures} for an illustration of our heterostructures
\cite{Samthesis}.

\begin{figure}[!ht]
\centering\includegraphics[width=4.5cm]{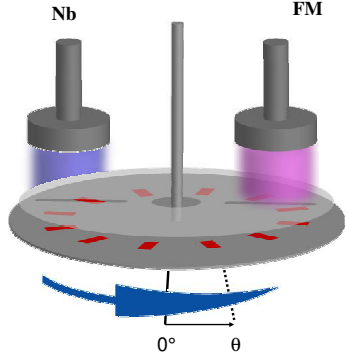} \caption{An
illustration of the sputtering process: the ferromagnetic interlayer
thickness is varied as a function of the angle $\theta$ from the
pre-sputter position. \label{sputter}}
\end{figure}
The heterostructures employed in this chapter, have been deposited by
d.c. magnetron sputtering. In a single deposition run,
multiple silicon substrates were placed on a rotating holder which
passed in turn under three magnetrons: Nb, Cu and the ferromagnetic
layer (Fe, Co, Py or Ni). The speed of rotation was controlled by a
computer operated stepper motor with a precision angle of better
than 3.6$^{\circ}$ and each sample was separated by an angle of at
least 10$^\circ$. Prior to the deposition each target material was
calibrated and the deposition rates were estimated with an Atomic
Force Microscopy (see table \ref{table} for a summary of the
deposition parameters for all target materials presented in this
chapter). In the case of the ferromagnetic materials (Co, Py, Fe and
Ni) the rates of deposition, and hence $t_{\rm F}$, were obtained by
varying the speed of each single chip which moves under the
ferromagnetic target while maintaining constant power to the
magnetron targets and Ar pressure. This was achieved by knowing the
chip position relative to the target material ($\theta$) and by
programming the rotating flange such that a linear variation of
ferromagnetic thickness with $\theta$, $d(t_F)/d\theta$, was
achieved. $t_F$ is inversely proportional to the speed of
deposition, $t_F \propto 1/V_t$, and hence it can be shown that in
order to achieve a linear variation of $t_F(\theta)$ one programmes
the instantaneous speed, at position $\theta$ and time $t$ seconds
(i.e. $V(\theta)_t$), of deposition according to
\begin{equation}
\ V(\theta)_t = \frac{V_i V_f}{V_f - V_i} \Bigg( \frac{V_i V_f}{V_f
- V_i}-\frac{\theta}{2 \pi} \Bigg ) ^{-1}, \label{rotation}
\end{equation}
where $V_i$ is the initial speed and $V_f$ is the final speed in
units of rpm. This method of varying $t_F$ guaranteed, in all
cases, that the interfaces between each layer were prepared under
the same conditions while providing precise control of the F layers
\cite{pijunctionPRB}. Fig. \ref{sputter} shows an illustration of
the sputtering process. For each run, simultaneously, our
heterostructures have been deposited on $5\times5$ mm$^2$ and
$5\times10$ mm$^2$ SiO$_2$ substrates. The first ones have been used
for the magnetic measurements, the second ones for the realization
of Josephson junctions.

\begin{table}
\centering
\caption{\textbf{\label{table}A summary of the deposition
parameters for all materials sputtered. $t_F$ refers to the
expected film thickness.}}\bigskip
\begin{tabular}{c c c c c} \hline \hline
Target material & Rate & Power & Speed range & t$_F$ \\
 & ($nm/W$ at 1 rpm) & (W) & (rpm) & ($\pm$0.2 nm) \\
\hline
Nb & 6.89 & 90 & .028 & 250\\
Cu & 4.39 & 30 & 0.22 & 20 \\
Ni & 2.12 & 40 & 0.20-4.2 & 0.5 - 10.5\\
Co & 1.8 & 40 & 0.36-3.6 & 0.5 - 5.0\\
Py & 1.64 & 40 & 0.33-3.3 & 0.5 - 5.0\\
Fe & 2.0 & 40 & 0.22-2.2& 0.5 -5.0\\
\hline \hline
\end{tabular}
\end{table}

\subsection{X-ray Measurements}
\begin{figure}[!ht]
\includegraphics[width=9.5cm]{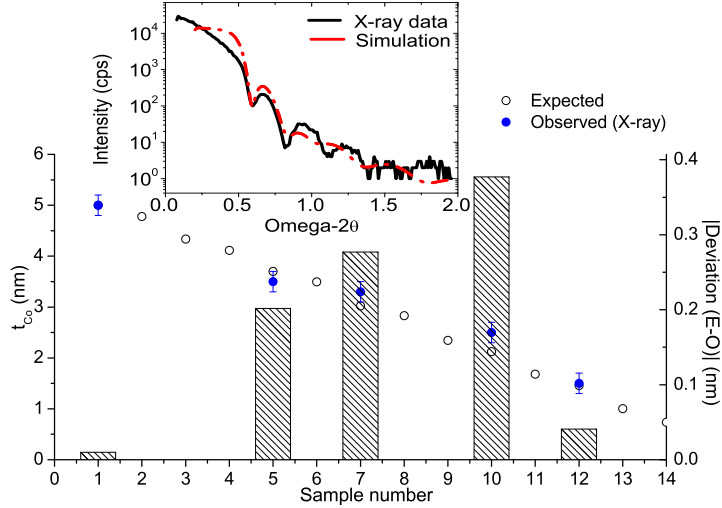}
\centering\caption{$t_{{\rm Co}(observed)}$ plotted with $t_{{\rm
Co}(expected)}$ extracted from simulation data, in the inset: low
angle x-ray data plotted with the equivalent simulation for sample 1
where $t_{{\rm Co}(expected)}=$5 nm and $t_{{\rm
Co}(observed)}=$5$\pm$0.2 nm. The mean deviation is estimated to be
$\pm$0.2 nm.\label{xray}}
\end{figure}

To confirm our precise control over the ferromagnetic thickness
variation we performed low angle X-ray reflectivity of a set of
calibration Nb/Co/Nb thin films where the Nb layers had a thickness
of $5$ nm and the Co barrier thickness was varied from $0.5$ nm to
$5.0$ nm \cite{pijunctionPRB}. A series of low angle X-ray scans
were made and the thickness of the Co layer ($t_{{\rm
Co}(observed)}$) was extracted by fitting the period of the Kiessig
fringes using a simulation package. It was found that our expected
Co barrier thicknesses, $t_{{\rm Co}(expected)}$, was well
correlated with $t_{{\rm Co}(observed)}$ with a mean deviation of
$\pm$0.2 nm. In the inset of Fig. \ref{xray} we show an example of
the low angle x-ray data plotted with the equivalent simulation data
where $t_{{\rm Co}(observed)}$ was extracted, while in Fig.
\ref{xray} we report a comparison of $t_{{\rm Co}(observed)}$ with
$t_{{\rm Co}(expected)}$ and its deviation.

\section{Nanoscale Device Process}

\begin{figure}[!ht]
\centering
\includegraphics[width=7cm]{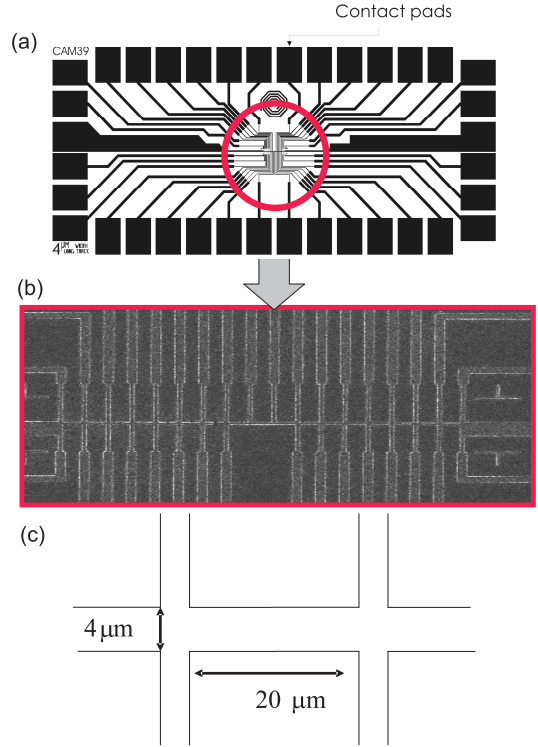}
  \centering \caption{(a)``Cam 39'' mask design for a $10\times5$ mm$^2$ chip used in this chapter, showing contact pads
for wirebonding. (b) Central detail: on each chip we realize 14
possible junctions with thinnest track widths of 4$\rm \mu m$ (c).
}\label{mask}
\end{figure}

We can summarize the realization of the Josephson junctions in three
different steps: (i) patterning the films using optical lithography
to define the micron scale tracks and the contact pads. Our mask
permits etching of at least 14 devices, which allows us to measure
numerous devices and to derive good estimates of important
parameters, like, for example, characteristic voltage (see Fig.
\ref{mask} to see an illustration of the mask); (ii) broad beam Ar
ion milling ($3$ mAcm$^{-2}$, $500$ V beam) to remove unwanted
material from around the mask pattern, thus leaving $4 \mu$m tracks
for subsequent FIB work; (iii) FIB etching of micron scale tracks to
achieve vertical transport
\cite{pijunctionPRL,IronEPJ,pijunctionPRB}.

\begin{figure}[!ht]
\includegraphics[width= 8cm]{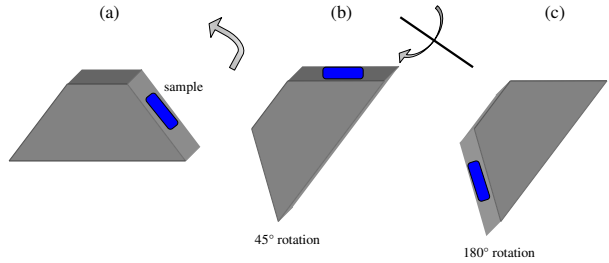}
\centering \caption[FIB sample holder]{\label{stage}Schematic of the
sample holder used in FIB (a) showing two axes of rotation,
$45^{\circ}$, to horizontal milling (b) and $180^{\circ}$ to
vertical milling.}
\end{figure}

In particular to realize our devices we have used a
three-dimensional
 technique \cite{Kim}. The wedge holder used in this chapter was designed by D.-J. Kang
 and it is schematically shown in Fig. \ref{stage}; it is
 constituted by three sample lodgings, one in horizontal and
 two at $45^\circ$. Once loaded the sample on the $45^\circ$ lodging,
 we can rotate the stage of the FIB at $45^\circ$, in this way the
 beam is perpendicular to the surface of the sample, and the first cut is done;
  then the sample holder is rotated of
 $180^\circ$ around an axis normal to the sample stage, to permit
 the vertical cut to be done (see Fig. \ref{stage}). This setup
 allows to load two samples for each run.

\begin{figure}[!ht]
\centering\includegraphics[width=5 cm]{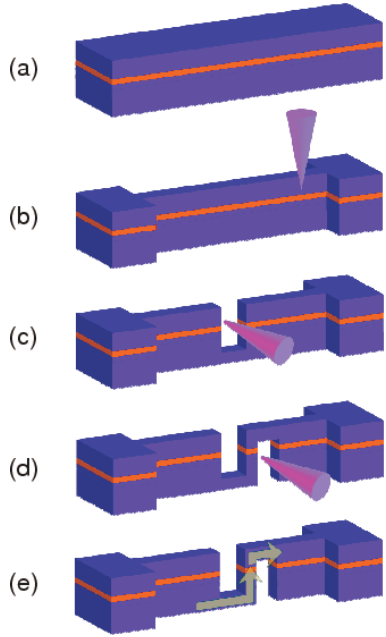} \centering
\caption{\label{FIBcut}FIB procedure for device fabrication: the
initial trilayer after the photolithography and ion-milling (a) is
cut with 150 pA beam (b) and then with 11 pA beam (c-d). (e) At the
end two side cuts are realized to create the final device structure
with a device area in the range of $0.2-1$ $\mu$m$^2$ achieving
vertical transport of the current. (Figure adapted from ref.
\cite{Blamire_Eucas}).}
\end{figure}

\begin{figure}[!ht]
\centering\includegraphics[width=7 cm]{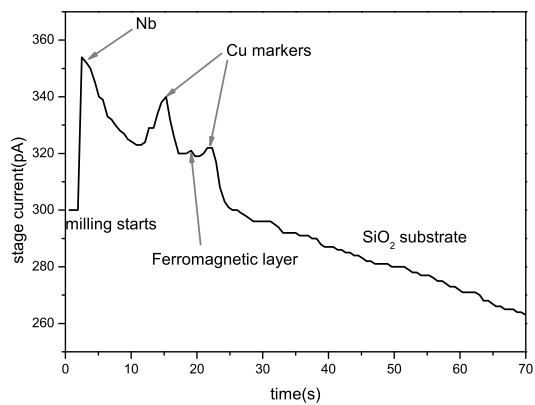} \caption[FIB sample
holder]{\label{EPD} End Point Detector. The graph shows the stage
current as a function of the milling time, we can distingue the Nb
layer, the two Cu layers, the Ferromagnetic barrier and when the
current decreases we have reached the substrate.}
\end{figure}

\begin{figure}[!ht]
\centering \includegraphics[width=11cm]{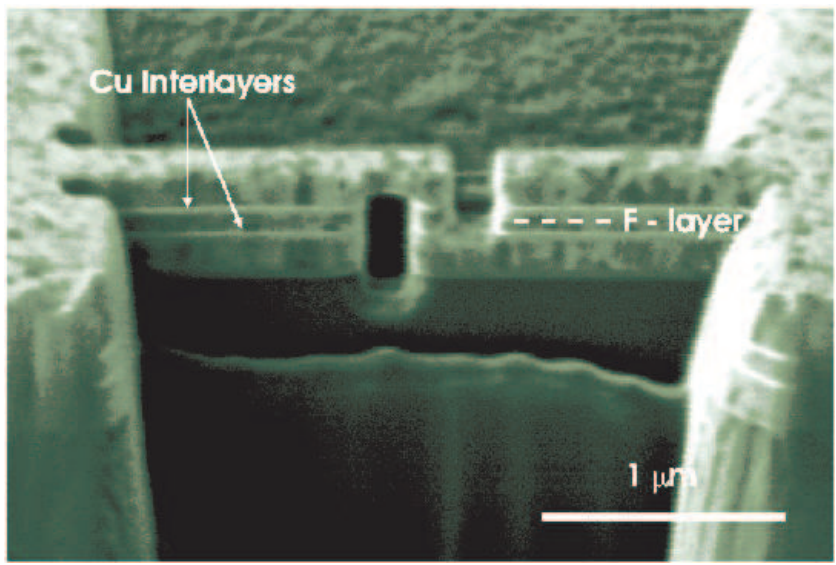}\\
  \caption{FIB image of a Nb/Cu/Co/Cu/Nb device, the two light gray lines are the Cu markers.}\label{junction}
\end{figure}

The fabrication procedure is shown in Fig. \ref{FIBcut}
\cite{Chrisnano}.  A first box with area $4 \times 2 {\rm \mu m^2}$
is milled with $150$ pA to realize tracks of  about $700$ nm (b).
The time of milling is about $1-3$ minutes, and this milling can be
calibrated using the stage-current/end-point detection measurement.
Fig. \ref{EPD} shows how the milling of different layers can be
distinguished, the first peak corresponds to the Nb, the two small
peaks correspond to the two Cu layers, and finally the intensity
decreases approaching to the SiO$_2$ substrate.
 The sidewalls of the narrowed track are then cleaned with a beam
current of  $11$ pA. This removes excessive gallium implantation
from the larger beam size of the higher beam currents. The cleaning
takes $\sim$ $25$-$30$ seconds per device. The track width is then
$\le 500$ nm. The sample is then tilted to $\theta = 90^{\circ}$,
and the two cuts are made with a beam current of $11$ pA to give the
final device (Fig. \ref{FIBcut} (d)) with a device area in the range
of $0.2-1$ $\mu$m$^2$. This technique permits to achieve vertical
transport of the current: in Fig. \ref{FIBcut} (e) we show a
schematic picture of the current path.  In Fig. \ref{junction} the
final FIB image of a Nb/Cu/Co/Cu/Nb device is presented.

\section{Magnetic Measurements}
In this section we report magnetic measurements of Nb Josephson
junctions with strongly ferromagnetic (F) barriers: Ni,
Ni$_{80}$Fe$_{20}$ (Py), Co and Fe. From measurements of the
magnetization saturation ($M_S$) as a function of the F thickness,
our heterostructures have shown a magnetic dead layer ranging
between $0.5$ nm and $1.7$ nm. Then we give an estimation of the
Curie temperature of the ferromagnetic layer.

\subsection{Measurement of the Magnetic Dead Layer}

 In this section we explain the magnetic properties of
 Nb-Ni-Nb, Nb-Py-Nb, Nb-Co-Nb and Nb-Fe-Nb heterostructures as a function of F
 thickness
 \cite{pijunctionPRL,pijunctionPRB,pijunctionIEEE,IronEPJ}.

\begin{figure}[!ht]
\centering
\includegraphics[width=7.5cm]{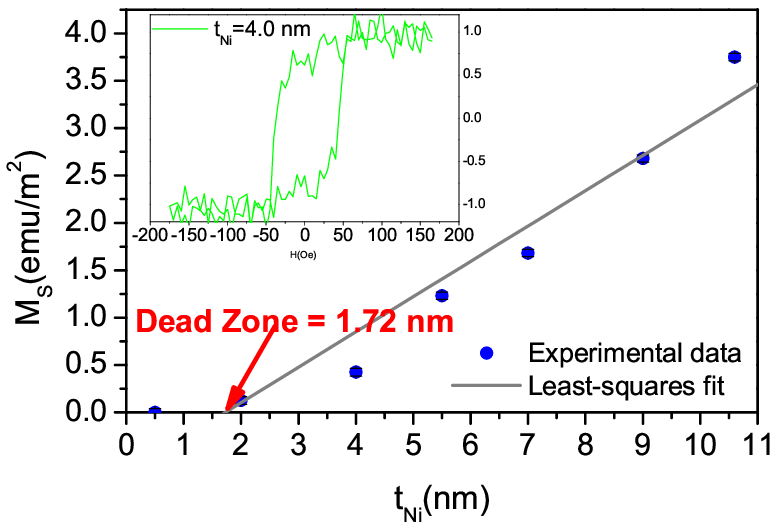}
\caption{Saturation magnetization versus thickness of the Ni
barrier, inset: hysteresis loop for $t_{\rm Ni}=4$ nm.\label{Nimag}}
\end{figure}

\begin{figure}[!ht]
\centering
\includegraphics[width=12cm]{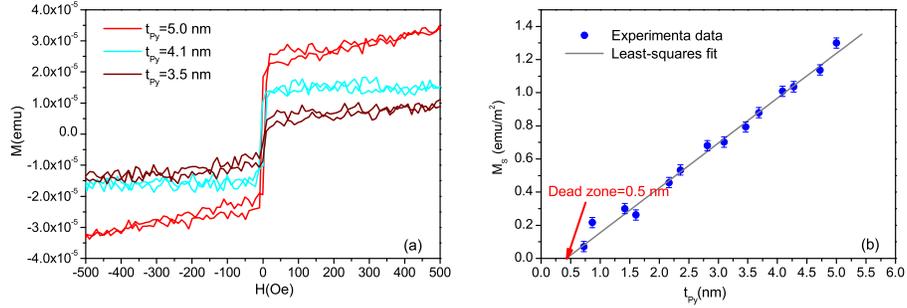}
\caption{(a) Hysteresis loops for different Py thicknesses. (b)
Saturation magnetization versus thickness of the Py barrier.
\label{Pymag}}
\end{figure}

\begin{figure}[!ht]
\centering
\includegraphics[width=12cm]{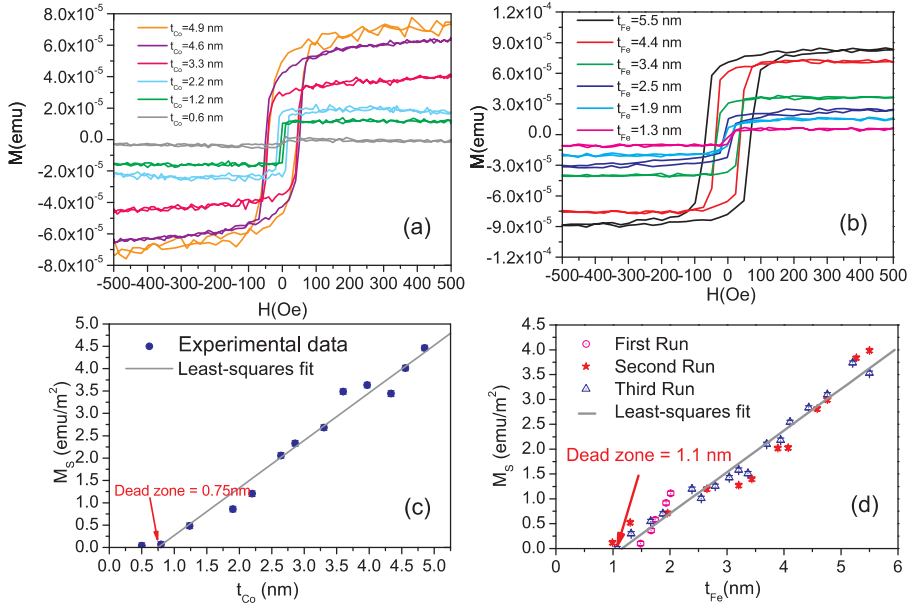}
\caption{Hysteresis loops for different Co (a) and Fe (b)
thicknesses. Saturation magnetization versus thickness of the Co (c)
and Fe (d) barrier. From the linear fit (gray line) the magnetic
dead layer is extrapolated. \label{Comag}}
\end{figure}

To this aim we have studied, using a VSM at room temperature, the
hysteresis loop of our heterostructures in order to follow the
evolution of the magnetization as a function of the applied magnetic
field. In Fig. \ref{Nimag} (inset),
 Fig. \ref{Pymag} (a), \ref{Comag} (a) and (b), we show a collection of
hysteresis loops for different thicknesses of the F barrier. We
notice that both the $M_S$ and the width of the hysteresis loop, as
expected, decrease with decreasing F layer thickness, and the
ferromagnetic order disappears when the F barrier goes to zero.
Furthermore from the hysteresis loops we have measured the
saturation magnetization as a function of the F barrier to
extrapolate the magnetic dead layer.

The presence of a magnetic dead layer has been reported
 in other studies of S/F/S
heterostructures \cite{cbellprb05,Born2006,pijunctionPRB} and it can
be explained as a loss of magnetic moment of the heterostructures.
The magnetic dead layer can be due to numerous factors, as, for
example, lattice mismatch at the Nb-F interface resulting in a
reduction in the ferromagnetic atomic volume \cite{Kitada} and
crystal structure which leads to a reduction in the exchange
interaction between neighboring atoms. This loss in exchange
interaction manifests itself as a loss in magnetic moment due to a
collapse in the regular arrangement of electron spin and magnetic
moment and can lead to the suppression in $T_{\rm Curie}$ and
$E_{ex}$ \cite{JAarts, Renjun,Qunwen,Pick}. Another factor can be
the inter-diffusion of the ferromagnetic atoms into the Nb. Like in
the case of lattice mismatch this would result in a breakdown of
the crystal structure at the interface leading to a reduction in the
exchange interaction. The knowledge of the magnetic dead layer is a
crucial point within the implementation of $\pi$-technology, to
guarantee the reproducibility and the control of the devices.
Extrapolating with a linear fit the saturation magnetization as a
function of F thickness, we have obtained the estimation of the
magnetic dead layer: $\sim$$1.7$ nm for Ni (see Fig. \ref{Nimag}),
$\sim 0.5$ nm for Py (see Fig. \ref{Pymag} (b)) and $\sim 0.75$ nm
for Co (see Fig. \ref{Comag} (c)).

\begin{figure}[!ht]
\centering
\includegraphics[width=8.5cm]{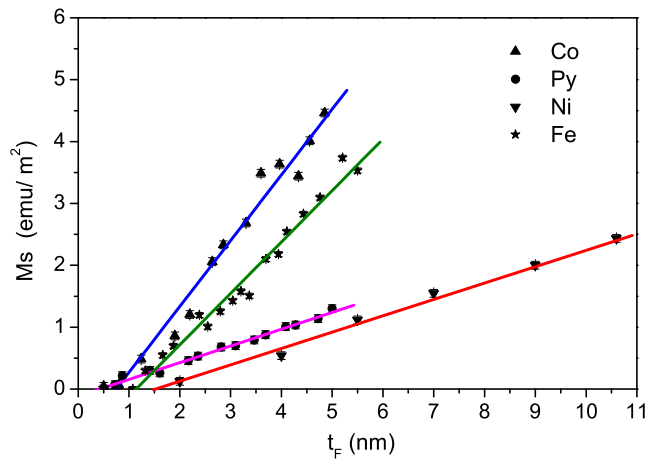}
\caption{Saturation magnetization \emph{vs} Ni, Py, Co and Fe
thickness at T$=300$K. \label{deadlayer}}
\end{figure}
In the case of Fe the saturation magnetization was measured for
three different deposition runs (see Fig. \ref{Comag} (d)). For each
deposition we have obtained similar saturation magnetization and by
extrapolating the least-squares fit of these data we have estimated
a Fe magnetic dead layer of  $\sim 1.1$ nm \cite{IronEPJ}.
 In Fig. \ref{deadlayer} we summarize the magnetic moment \emph{vs} thickness
for Ni, Py, Co and Fe. From these data we can extrapolate the slope
of $M_{S}$ \emph{vs} $t_{\rm F}$. From the theoretical model
\cite{Slater,Pauling} the predicted slopes are: $\simeq 0.60$
emu/cm$^3$ for Ni, $\simeq 0.52$ emu/cm$^3$ for Py, $\simeq 1.42$
emu/cm$^3$ for Co and $\simeq 2.6$ emu/cm$^3$
 for Fe. On the other hand from our
experimental data the slopes of $M_{S}$ \emph{vs} $t_{\rm F}$
for these ferromagnetic materials are suppressed from these expected
bulk values: $0.27$ emu/cm$^3$ for Ni, $0.27$ emu/cm$^3$ for Py,
$1.0$ emu/cm$^3$ for Co and $0.83$ emu/cm$^3$ for Fe. We can argue
that, in our case, we are not considering bulk materials, as
reported from Slater and Pauling \cite{Slater,Pauling}, instead our systems are
constituted by F/S sandwiches, so the presence of a superconducting
layer, Nb, can induce a weakening of the ferromagnetic properties of
the F layer and a possible diffusion of the Nb into the F barrier.

\subsection{Calculation of the Curie temperature}
\begin{figure}[!ht]
\centering\includegraphics[width=9cm]{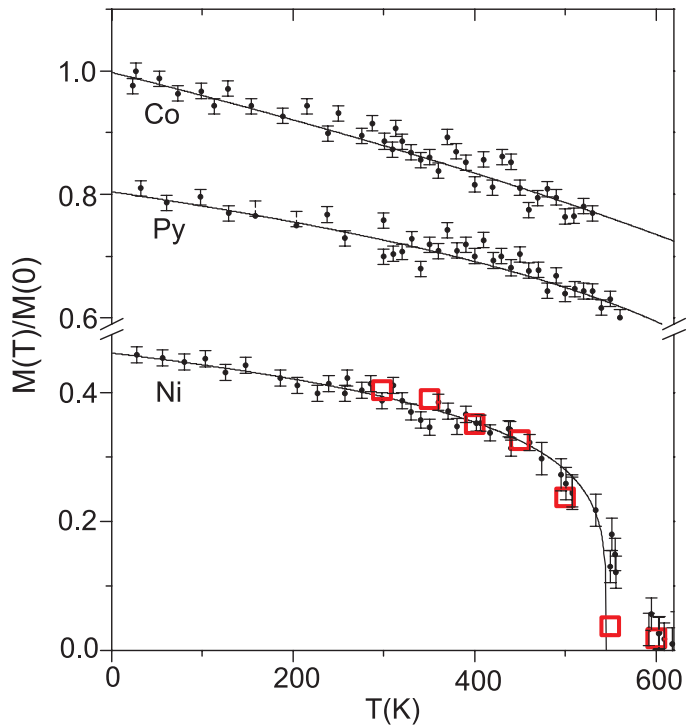} \caption{Thermal
variation of the magnetization for Co, Py and Ni (black dots) with
the best fitting curves to extrapolate the Curie temperature of Ni
(571K), Py (800 K) and Co (1200K)\label{nimag2}. Red squares are
cooling data for Ni.}
\end{figure}
We have measured the thermal variation of the saturation
magnetization, $M(T)$, of Co, Ni, and Py when sandwiched between
thick Nb layers \cite{pijunctionPRB}. To be certain, $M(T)$ with
temperature was not weakened by a thermally activated diffusion of
ferromagnetic atoms into Nb, or vice versa; at the interface we
measured both the magnetization when warming and cooling. The
warming and cooling data agreed for all three barrier systems up to
a temperature of 620 K. Above this temperature, $M_S$ was found to
drop by virtue of thermally activated diffusion. We have modeled the
warming and cooling data of $Ms$ $(10 {\rm K}<T<620 {\rm K})$ with
the following formula
$$M(T)/M(0)=(1-T/T_{\rm Curie})^{\beta},$$
where $M(0)$ is the saturation magnetization at absolute 0 K, $T$ is
the measuring temperature, and $\beta$ and $T_{\rm Curie}$ are fitting
parameters. This gives $T_{\rm Curie}$ values of 1200 K for Co, 571 K
for Ni, and 800 K for Py (see Fig.\ref{nimag2}) in agreement with
the bulk values. Data for Ni are the most reliable because we have a
full data set. However, in any case these measurements provide that,
in our metallic systems, interdiffusion at the ferromagnetic surface
cannot be ruled out because the ferromagnetic layer is known to form
a variety of magnetic and nonmagnetic alloys with Nb.

\section{Transport Measurements}
In this section we present the $I(V)$  \emph{vs} $V$ curves for the
Nb/F/Nb Josephson junctions varying the thickness of the F barrier.
All these materials show multiple oscillations of the Josephson critical current
with barrier thickness implying repeated $0$-$\pi$ phase-transitions
in the superconducting order parameter. The critical current
oscillations have been modeled with the clean and dirty limit
theoretical models, from this analysis we have extrapolated the
exchange energy and the Fermi velocity of the ferromagnetic barrier.

\subsection{Experimental Data: Critical Current Oscillations}
\begin{figure}[!ht]
\centering
\includegraphics[width=8.5cm]{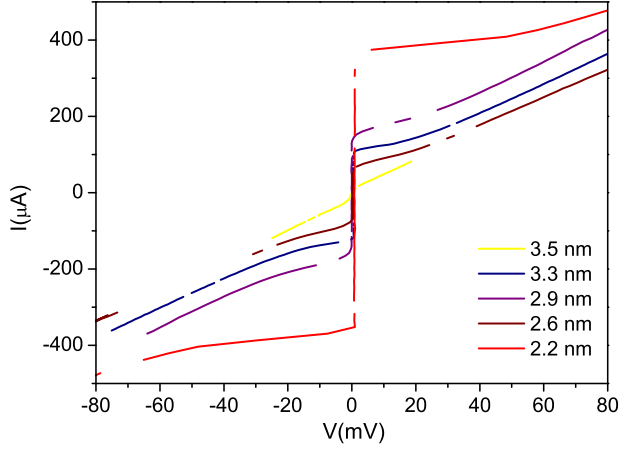}
\caption{$I-V$ characteristics for Nb/Co/Nb Josephson junctions with
different Co thickness.\label{Codiff_chips}}
\end{figure}
 For several junctions on
each chip we have measured the current versus voltage from which the
critical current ($I_c$) and the normal resistance ($R_N$) of the
Josephson junction have been extrapolated. In Fig.
\ref{Codiff_chips} we show an example of $I$ \emph{vs} $V$ curves
for different Co thicknesses, we notice that for each curve we have
roughly the same resistance, that means the same area of the
Josephson junctions. This is an indication of the good control on
the fabrication of the devices with the FIB.

\begin{figure}[!ht]
\centering
\includegraphics[width=8.5cm]{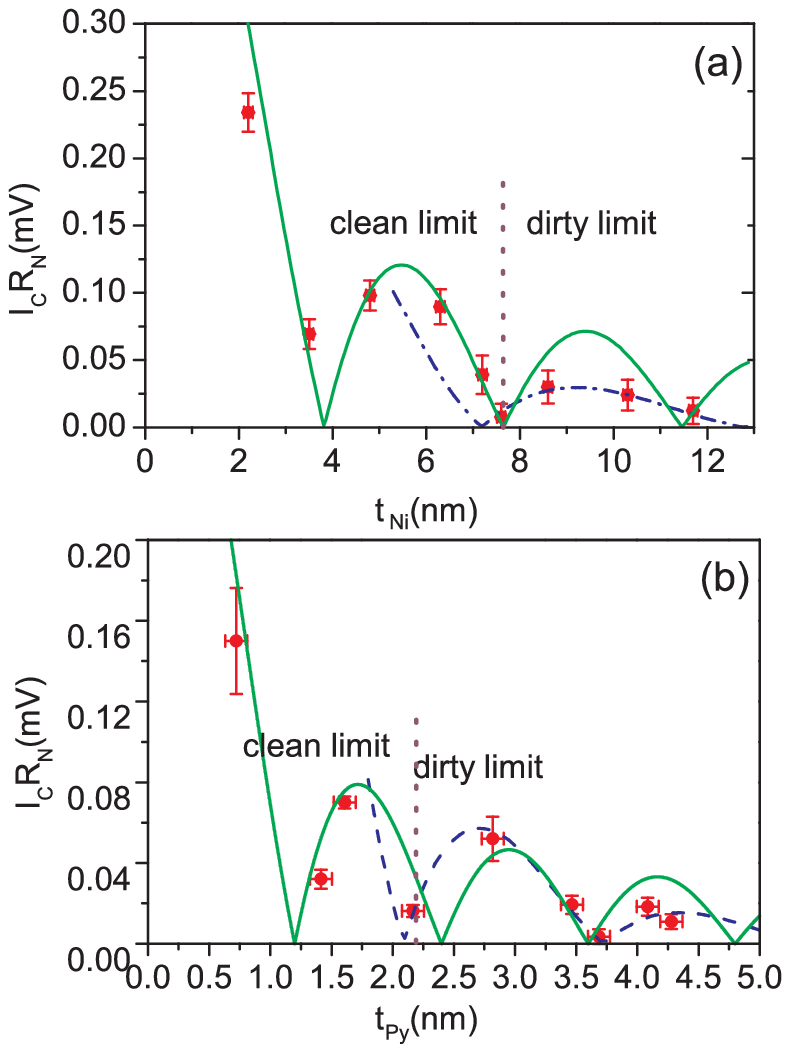}
\caption{Oscillations of the $I_CR_N$ product as a function of the
thickness of the Ni (a) and Py (b) barrier. The solid green line is
the theoretical fit in agreement with Eq. \ref{cleanlimit}, dash-dot
blue line is referred to Eq. \ref{dirtylimit}. \label{NiPyOscill}}
\end{figure}

\begin{figure}[!ht]
\centering
\includegraphics[width=8.5cm]{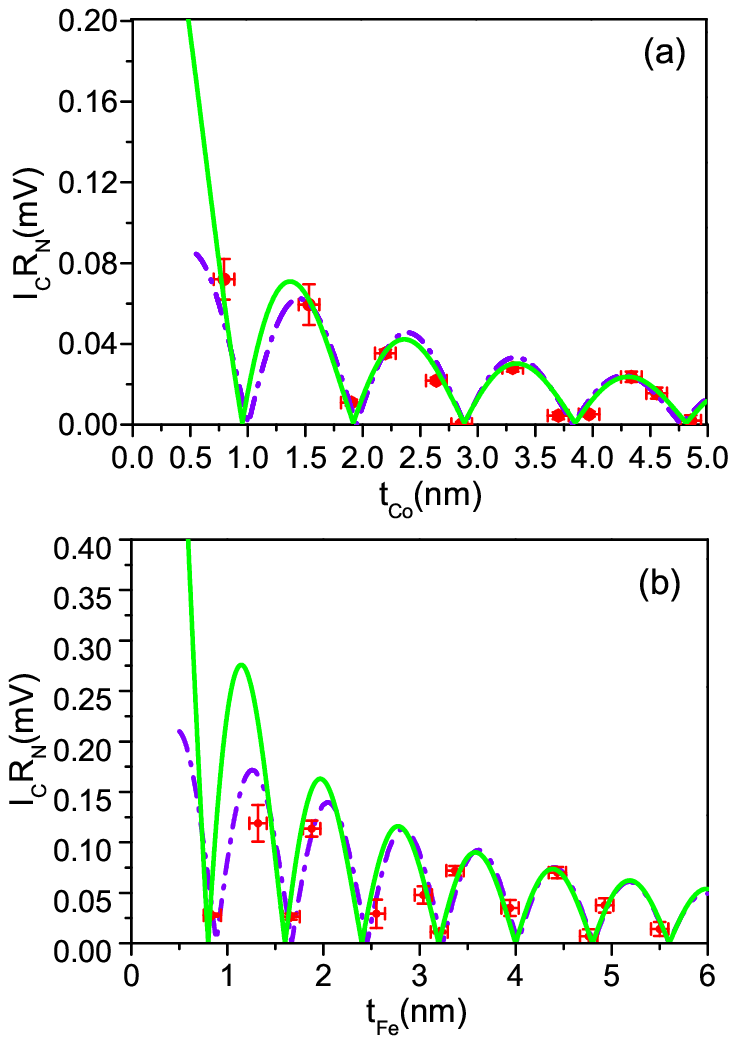}
\caption{$I_cR_N$ \emph{vs} $t_{\rm Co}$ (a) and $t_{\rm Fe}$ (b)
with the best fitting theoretical model in agreement with Eq.
\ref{general}, dash-dot violet line, and Eq. \ref{cleanlimit} solid
green line.\label{CoFe_Oscill}}
\end{figure}

The $I_cR_N$ products, as a function of F barrier thickness, exhibit
a decaying, oscillatory behavior, in agreement with the theoretical
predictions. The oscillations of $I_{c}R_{N}$ as a function of Ni,
Py, Co and Fe thicknesses at $4.2$ K are shown in Figs.
\ref{NiPyOscill} (a)-(b) and Figs. \ref{CoFe_Oscill} (a)-(b)
\cite{pijunctionPRL,IronEPJ,pijunctionPRB}.

In the case of Py, the clean limit model, Eq. \ref{cleanlimit},
closely matches the experimental data up to a thickness of $\simeq
2$ nm, while in the case of Ni the oscillations are explained by the
clean limit theory up to $\simeq 7$ nm. Above these values a better
fit is obtained using a formula for a diffusive and high $E_{ex}$
ferromagnet, Eq. \ref{dirtylimit}. For Eq. \ref{cleanlimit} the
fitting parameters are the $E_{ex}$ and the Fermi velocity $v_F$. In
the case of Eq. \ref{dirtylimit} the fitting parameters are: $v_F$,
the mean free path $L$, the superconducting energy gap $\Delta$ and
$E_{ex}$. But $L$ and $\Delta$ are not free parameters, because they
are fixed by the theoretical predictions. In this way the only free
parameters are, as for the clean limit equation, the Fermi velocity
and the exchange energy. Using Eq. \ref{dirtylimit} we obtain the
best fitting values  for Ni: $v_F ({\rm Ni}) = 2.8\times 10^5$ m/s
and $L_{\rm Ni}\simeq 7$ nm; while for Py $v_F ({\rm Py}) =
2.2\times 10^5$ m/s, $L_{\rm Py}\simeq 2.3$ nm and $\Delta = 1.3$
meV. These values are consistent with the ones used in Eq.
\ref{cleanlimit} and elsewhere \cite{Blum,cbellprb05}; while for the
exchange energy we estimate $E_{ex}({\rm Ni})\simeq 80$ and
$E_{ex}({\rm Py})\simeq 201$ meV. $E_{ex}({\rm Ni})$ is close to
other reported values by photoemission experiments \cite{Heinmann},
but smaller than that reported by other authors \cite{Blum}. The
smaller than expected $E_{ex}({\rm Ni})$ is a consequence of
impurities and possibly interdiffusion of Ni into Nb. Anyway, from
the magnetic measurements the extrapolated value of the $T_{\rm
Curie}$ provides evidence that our Ni is of acceptable quality. The
$E_{ex}({\rm Py})$ is consistent with the expected value and is
approximately twice that measured in Nb/Py/Nb junctions deposited
with epitaxial barriers where $E_{ex}\simeq 95$ meV
\cite{cbellprb05}.

On the other hand for Co and Fe the thicknesses range all in the
clean limit, so the experimental data have been modeled with the
general formula, Eq. \ref{general}, and $\xi_1$ and $\xi_2$ are the
two fitting parameters. We can see that the experimental data are in
good agreement with the theoretical model.

In particular for the Co data, from the theoretical fit shown in
Fig. \ref{CoFe_Oscill}(a) we find that the period of oscillations is
$T=1.91$ nm, hence $\xi_2 \sim 0.30$ nm and $\xi_1 \sim 3.0$ nm. In
the clean limit, $\xi_2= v_F \hbar/2 E_{ex}$. In this way, known
$\xi_2$ and estimating the Fermi velocity of being $v_F=2.8 \times
10^5$ m/s, as reported in literature, we can calculate the exchange
energy of the Co: $E_{ex}= \hbar v_F/2T\approx 309 {\ \rm meV}$.

For Fe, from the theoretical fit we obtain $\xi_1=3.8$ nm and
$\xi_2=0.25$ nm. So the period of the oscillations is $T=1.6$ nm.
Known $\xi_2$ and $v_F=1.98 \times 10^5$ m/s \cite{literature}, we
can calculate the exchange energy of the Iron: $E_{ex}= \hbar
v_F/2T\approx 256 {\ \rm meV}$.

To confirm that our oscillations are all in the clean limit (meaning
that the considered Co and Fe thickness are always smaller than the
Co and Fe mean free path), we have modeled our data  with the
simplified formula which holds only in this limit, Eq.
\ref{cleanlimit}, where in this case $E_{ex}$ and  $v_F$ are the two
fitting parameters. From the theoretical fit we obtain
$E_{ex}({\rm Co})=309$ meV, $v_F({\rm Co})=2.8 \times 10^{5}$ m/s and
$E_{ex}({\rm Fe})=256$ meV, $v_F({\rm Fe})=1.98 \times 10^{5}$ m/s. We remark
that the best fits are obtained with exactly the same values as
previously reported from Eq. \ref{general}. Both models provide
excellent fits to our experimental data showing multiple
oscillations of the critical current in a tiny (nano-scale) range of
thicknesses of the Co and Fe barrier.

\subsection{Estimation of the Mean Free Path}

\begin{figure}[!ht]
\centering
\includegraphics[width=8.5cm]{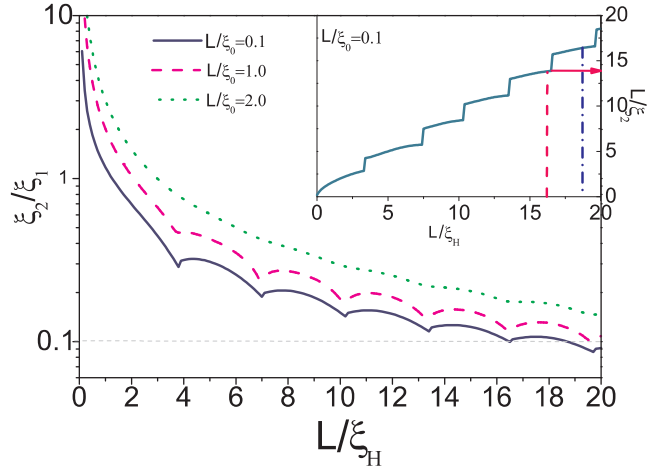}
\caption{Estimation of Co mean free path. The dependence of $\xi_2
/\xi_1$ with inverse magnetic length, $L/\xi_H$, calculated for
different ratios of $L/\xi_0$. Inset: inverse decay length,
$L/\xi_2=f(L/\xi_0)$ for when $L/\xi_H \simeq 0.1$
\label{paperBorn}}
\end{figure}

\begin{figure}[!ht]
\centering
\includegraphics[width=8.5cm]{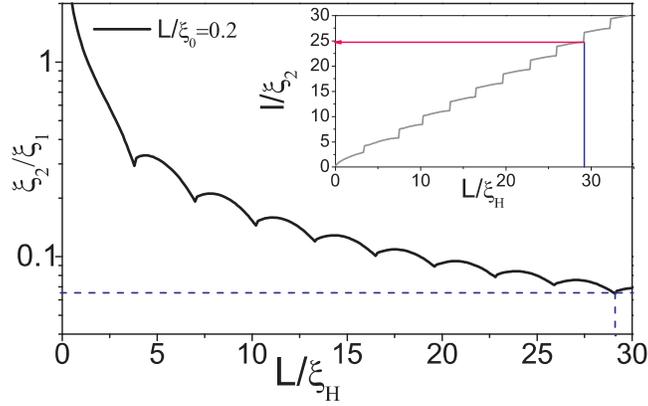}
\caption{Estimation of Fe mean free path. $\xi_2 /\xi_1$ \emph{vs}
$L/\xi_H$, calculated for different ratios of $L/\xi_0=0.2$.Inset:
$L/\xi_H$ \emph{vs} $L/\xi_H$ to estimate the mean free path, $L$.
\label{Fe_paperBorn}}
\end{figure}

With a simplified model that is obtained solving the linear
Eilenberger equations \cite{Gusakova} we can estimate the mean free
path for Co and Fe. The general formula is:
\begin{equation}
\ \tanh \frac{L}{\xi_{eff}}=\frac{\xi_{eff}^{-1}}{\xi_0^{-1} +
L^{-1} + i \xi_H^{-1}} \label{Born}
\end{equation}
where $\xi_{eff}$ is the effective decay length given by
$\xi_{eff}^{-1} = \xi_1^{-1} + i \xi_2^{-1}$, $\xi_o$ is the
Ginzburg-Landau coherence length and $\xi_H$ is a complex coherence
length. In the clean limit $1+L \xi_0^{-1} \gg \frac{1}{2}$max$\{ \ln
(1 + L \xi_0^{-1}), \ln(L \xi_H^{-1})\}$. The solution of Eq.
\ref{Born} gives
\begin{equation}
\ \xi^{-1}_1 = \xi_0^{-1} + L^{-1}, \xi_0=\frac{v_F \hbar}{2 \pi T_c
k_B}, \xi_2 = \xi_H = \frac{v_F \hbar}{2 E_{ex}}, \label{Eilen}
\end{equation}\label{born2}
and the numerical solution is shown in Fig. \ref{paperBorn} for Co
and in Fig. \ref{Fe_paperBorn} for Fe.

In the case of Co \cite{pijunctionPRL}, following this method we
find from Fig. \ref{paperBorn} that the experimental ratio
$\xi_2/\xi_1\simeq 0.1$ corresponds to two inverse magnetic lengths
of $L/\xi_H\simeq 16.5$ and $L/\xi_H\simeq 18.7$. By assuming
$L/\xi_0\simeq 0.1$ and for the estimated parameters $\xi_1\simeq 3$
nm and $\xi_2\simeq 0.3$ nm we obtained, from the inset in Fig.
\ref{paperBorn} a mean free path $L\simeq 5$ nm. Furthermore to
validate the determined mean free path of our Co thin film we have
estimated $L_{\rm Co}$ in a $50$ nm thick Co film by measuring its
resistivity as a function of temperature using the Van der Pauw
technique. The transport in Co thin films is dominated by
free-electron-like behavior \cite{Gurney} and hence the maximum mean
free path is estimated from $L=\hbar k_F / n_e e^2 \rho_b$, where
$n_e$ is the electron density for Co and is estimated from the
ordinary Hall effect to be $5.8 \times 10^{28}$ cm$^{-3}$ \cite{Gil}
and $\rho_b$ is the residual resistivity. The residual resistance
ratio ($(\rho_T + \rho_b)/\rho_b$) was measured to be $\simeq 1.41$
and hence we calculate $L_{\rm Co}$ at $4.2$ K to be $10$ nm. This
verifies that our assumption that for Co the oscillations are all in
the clean limit is well justified.

The same method can be done to estimate the mean free path for Fe
\cite{IronEPJ}. In this case the ratio $\xi_2/\xi_1 \simeq 0.06$, so
supposing $L/\xi_0$=$0.2$, we can extrapolate from the graphical
solution (Fig. \ref{Fe_paperBorn}) a value of $L/\xi_H \simeq 29$.
Considering the curve $L/\xi_2$ \emph{vs} $L/\xi_H$ (see inset in
Fig. \ref{Fe_paperBorn}) we obtain a value $L/\xi_2$ of about $25$
and we estimate the value of the mean free path of about $6$nm. With
this analysis we can remark that for Co and Fe the condition that
all the thicknesses are in the clean limit is unambiguously
fulfilled.

\subsection{Shapiro Steps and Fraunhofer Pattern}

\begin{figure}[!ht]
\centering
\includegraphics[width=12cm]{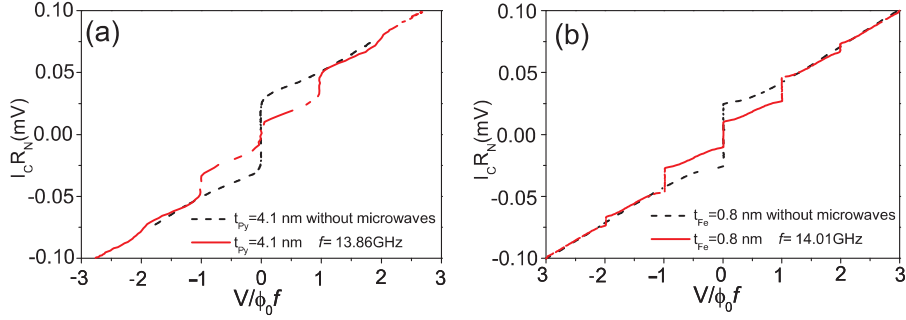}
\caption{A typical I $vs$ V curve of a  Josephson junction with 4.1
nm of Py (a) and 0.8 nm of Fe (b) (black dash line) and the
 integer Shapiro steps in the voltage-current curve with an excitation at
13.86 GHz (a) and 14.01 GHz (red line). $\phi_0=h/2e$
\label{shapiro}}
\end{figure}

 In Fig. \ref{shapiro} we show an example of typical $I$ \emph{vs} $V$ characteristics for a sample
 with a barrier of $4.1$ nm of  Py (a) \cite{pijunctionIEEE} and $0.8$ nm of Fe (b) \cite{IronEPJ}.

For these $I(V)$ curves we also show the effect of an applied
microwave field with an excitation at $f=13.86$ GHz and $f=14.01$
GHz, respectively. Constant voltage Shapiro steps appear due to the
synchronization of the Josephson oscillations on the applied
excitation \cite{Barone}. As expected the steps manifest at voltages
equal to integer multiples of $hf/2e$.

\begin{figure}[!ht]
\centering\includegraphics[width=9cm]{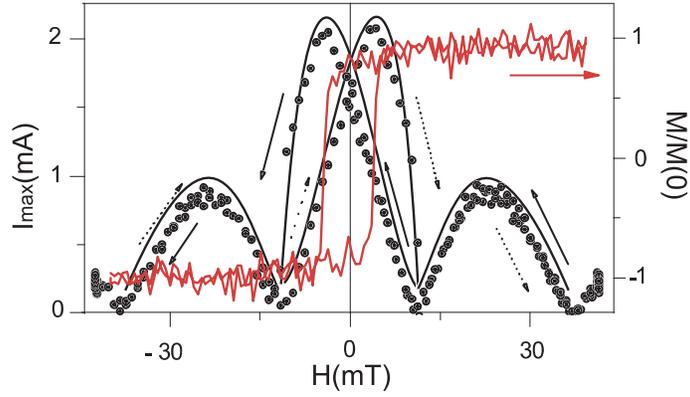}\\
  \caption{A typical Fraunhofer pattern for a device with 2.5 nm thick Fe barrier. The central peak is offset by 4.9 mT, which corresponds to
  a coercitive field of the Fe barrier, as confirmed by the hysteresis loop (red line).}\label{fraun}
\end{figure}
We have looked at the effect of a magnetic field on the maximum
supercurrent $I_{max}$ in our devices. We find that $I_{max}$
oscillates with applied magnetic field, giving rise to a Fraunhofer
pattern; however, we also find that $I_{max}$, which normally
corresponds to the central peak of a Fraunhofer pattern, is offset
from zero applied field to $\pm H_{offset}$, which is equal to $\pm
4.9$ mT. Fig. \ref{fraun} shows a typical Fraunhofer pattern for a
Fe barrier device with a barrier thickness of $\simeq 2.5$ nm. We
compared the variation of $I_{max}$ with applied field to the
magnetic hysteresis loop of the same film (measured prior to
patterning and device fabrication) at 20 K. The offset field is
found to correspond approximately to the coercive field of the
unpatterned film, which is $\pm H_{coercive}\simeq 4.3$ mT. The
central peak is shifted by the coercive field in each direction,
which is due to the changing magnetization of the ferromagnetic
barrier. The side peaks are not hysteretic and displaced by the
saturation moment of the barrier because the hysteresis loop is
saturated and the barrier moment is constant for both field sweep
directions. The coercive field and offset field in the Fraunhofer
pattern do not exactly agree: firstly, the coercive field is
approximated at 20 K; secondly, processing a device in a FIB
microscope is likely to harden the Fe magnetic domains by virtue of
Ga ion implantation. Similar results were measured for Co, Ni, and
Py.

\subsection{Half Integer Shapiro Steps Close to the 0-$\pi$ Transition}
\begin{figure}[!ht]
\centering
\includegraphics[width=12cm]{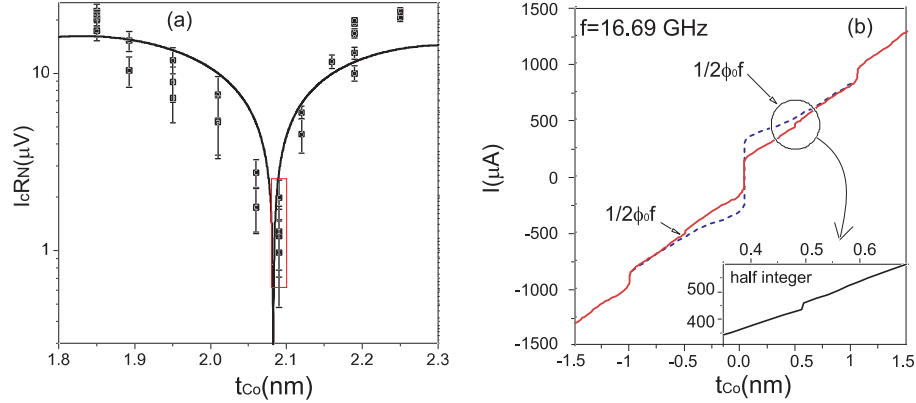}
\caption{(a) $I_cR_N$ versus $t_{\rm Co}$ compared to the fit of Eq.
\ref{cleanlimit} used as an eye guide only; (b) $I$ \emph{vs} $V$
for $t_{\rm Co}=2.1$ nm without (dashed blue line) and with (solid red line) microwaves at $f=16.69$
GHz; this device exhibits a sudden drop in differential resistance
at $V/n \phi_0f$ =0.5. The inset shows the details of the half
integer Shapiro step. \label{halfshapiro}}
\end{figure}

In this subsection, we present a study of a single transition from
the 0 to $\pi$ state in Co devices where $t_{\rm Co}$ thickness
varied from 1.8 to 2.5 nm \cite{pijunctionPRB}. The Co magnetic dead layer was estimated
to be 1.2 nm; although larger than the previously estimated dead layer of 0.8 nm   for the Co barriers, we
found that the bulk magnetizations for the two sample sets are
similar. Importantly, the magnetic data convincingly showed
incremental increases in magnetic moment with increasing Co
thickness. From current-voltage measurements, $I_c$ and $R_N$ were
extracted so that $I_cR_N$ could be determined and tracked as a
function of $t_{\rm Co}$. The characteristic voltage decreases to a
small voltage around a mean Co barrier thickness of 2.05 nm and then
increases, implying a change in phase of $\pi$. Each datum point  in Fig.
\ref{halfshapiro}(a) was
 measured, and the vertical error bars were derived
from a combination of estimating $I_c$ and $R_N$ from the
current-voltage curves and from a small noise contribution from the
current source. Particularly for those devices with the smallest
characteristic voltage, there was a considerable scatter in the obtained
$I_cR_N$ values. For an eye guide only, we have modeled the
transition with Eq. \ref{cleanlimit}, assuming $E_{ex}({\rm Co})=309$ meV. The curve is offset to fit the experimental data. The model
ignores any influence of a second harmonic term in the current-phase
relation.

In general, the current-phase relation is periodic in
$\phi$, the phase difference; however, recent theoretical and
experimental works \cite{Sellier} have looked at the possibility of observing
higher order harmonics, $I_s =I_{c1} \sin \phi +I_{c2}  \sin (2\phi)$,
where $I_{c2}\gg I_{c1}$ at the 0 to $\pi$ crossover and the second
harmonic dominates. When $I_s =I_{c2} \sin(2\phi)$, one expects both
integer and half-integer Shapiro steps in the current-voltage curves
at particular microwave frequencies. In our experiment \cite{pijunctionPRB}, by applying microwaves in the
13--17 GHz range to those devices near the transition, we have found
that the device with the smallest characteristic voltage and with a
Co barrier thickness of 2.1 nm exhibited current steps at both
half-integer $n=1/2$ and integer $n=1$ values of $V/n \phi_0 f$, as plotted in Fig. \ref{halfshapiro}(b).
These results imply that this device is close to a minimum
characteristic voltage and provides evidence for a second
harmonic in the current-phase relation; anyway further
investigations must be done before any conclusive remarks can be
made.

\subsection{Temperature Dependence of the $\boldsymbol{I_cR_N}$ Product}

\begin{figure}[!ht]
\centering\includegraphics[width=7cm]{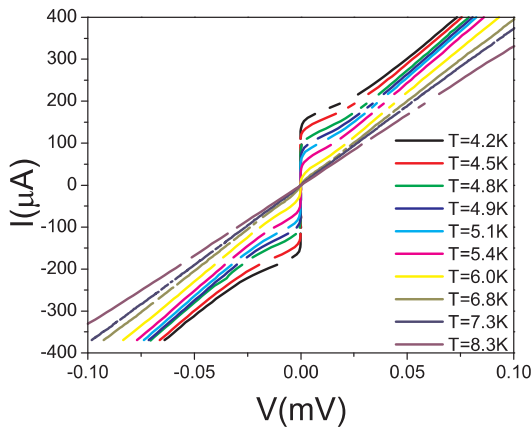}\\
  \caption{Temperature dependence of the Nb/Co/Nb Josephson junction with $2.2$ nm
  of Co.}\label{Cotemp}
\end{figure}

\begin{figure}[!ht]
\centering\includegraphics[width=12cm]{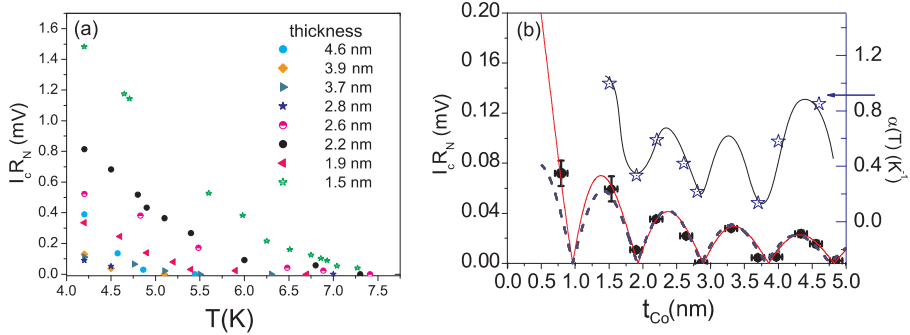}\\
  \caption{(a) $I_CR_N$ \emph{vs} $T$ for different $t_{\rm Co}$. (b) Oscillations of the decay
  rate $\alpha_T(T_0=4.2{\rm K})$ of $I_CR_N$, Eq.~\ref{alpha}, with the temperature [stars; the blue line is a guide to the eye],
  as compared with the oscillations of the $I_CR_N$ product itself in Nb/Co/Nb $\pi$ junctions.}\label{Cotemp3}
\end{figure}

\begin{figure}[!ht]
\centering\includegraphics[width=12cm]{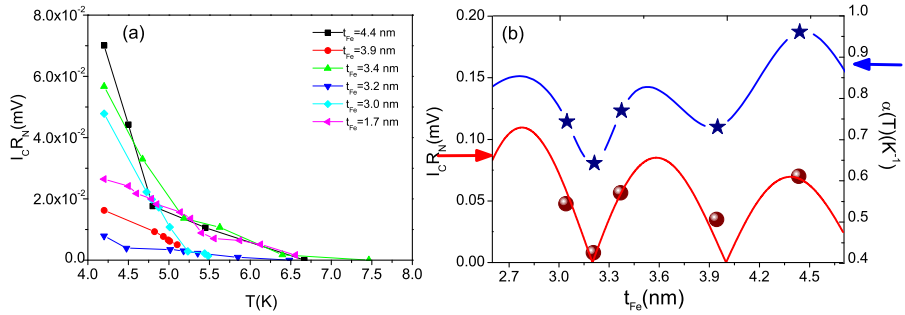}\\
  \caption{(a) Temperature dependence of the $I_CR_N$ for
  different thicknesses of the Fe layer. (b) Oscillations of the decay
  rate $\alpha_T(T_0=4.2{\rm K})$ of $I_CR_N$, Eq.~\ref{alpha}, with the temperature [stars; the blue line is a guide to the eye],
  as compared with the oscillations of the $I_CR_N$ product itself [spheres;
  the red line is the fit given by Eq.~\ref{general}] in a window of Fe thicknesses.}\label{tempesta}
\end{figure}

Further evidence of the oscillatory dependence of the characteristic
voltage is given by measurements of its thermal variation.  We
notice that, for each Co thickness, the Josephson junction
resistance remains approximately constant, slightly increasing when
approaching the critical temperature of Nb ($T_C = 9.2$K), as
expected; this is shown in Fig.\ref{Cotemp} for $t_{\rm Co}$=2.2 nm.
On the other hand the critical current, and hence the product
$I_CR_N$, quickly decreases with increasing $T$ and goes to zero at
the critical temperature of the Nb. This is shown explicitly in Fig.
\ref{Cotemp3} (a) for different Co thicknesses.


As for the Co data, in Fig. \ref{tempesta} (a) we show the
temperature dependence of the $I_CR_N$ product for different
thicknesses of the Fe barrier. We remark that also in this case the
$I_CR_N$ product, quickly decreases with increasing the
temperature \cite{IronEPJ}.

It is interesting to notice how the rate at which the critical
current drops to zero, exhibits a non-monotonic dependence on the Co
and Fe layer thickness. We can define the relative decay rate
$\alpha_T$ of the $I_CR_N$ product with the temperature as in
Ref. \cite{Blum}, namely
\begin{eqnarray}\label{alpha}
\alpha_T(T_0)&=&\left|\frac{{\rm d} \ln[I_CR_N(T)]}{{\rm
d}T}\right|_{T=T_0} \\ &=& \left|\frac{{\rm d} I_CR_N(T)}{{\rm
d}T}\right|_{T=T_0} \frac{1}{I_CR_N(T_0)}\,, \nonumber
\end{eqnarray}
calculated at $T_0=4.2$ K. In Fig. \ref{Cotemp3} (b) and in Fig.
\ref{tempesta} (b) we plot $\alpha_T$ as a function of $t_{\rm Co}$
and $t_{\rm Fe}$, respectively, for consecutive thicknesses; we
observe oscillations of $\alpha_T$ in phase with the oscillations of
the $I_CR_N$, in other words the smaller $I_CR_N$ decays to zero
with increasing $T$ more slowly than the greater $I_CR_N$. A
satisfactory explanation for this intriguing phenomenon awaits
further investigation.

\section{Conclusion}
In summary, we have fabricated and investigated Nb/F/Nb Josephson
junctions, with F standing for a strong ferromagnetic layer of Ni, Py, Co or Fe, varying the F barrier thickness. A magnetic
dead layer lower than $1.7$ nm is extrapolated by a linear
regression of the thickness dependence of the saturation
magnetization. From the $I(V)$ \emph{vs} $V$ curves we have measured
the Josephson critical current $I_C$ and the normal resistance
$R_N$ in order to follow the oscillations of the $I_CR_N$ product as
a function of the F thickness. In agreement with the theoretical
models for the clean and dirty limit, we have fitted the
experimental data estimating the exchange energy and the Fermi
velocity of the F barrier (see table \ref{summary} for a summary of
the $E_{ex}$ and $v_F$). Shapiro steps appear at integer multiples
of the applied voltage. Then, for different thickness of the Co and
Fe barrier, we have shown that the $I_CR_N$ product decreases
with increasing temperature, and in particular the decay rate
presents the same oscillatory behavior as the critical current.

With the results of this chapter, we have shown that we are able to
produce nano-structured Nb/F/Nb $\pi$-junctions with a high control
of the F thickness (within an accuracy of $0.2$ nm). The low
magnetic dead layer gives us the possibility to reproduce the
transport properties in our heterostructures with a small thickness
deviation. The estimated $E_{ex}$ is close to that of the bulk F,
which implies that our films are deposited cleanly with only a small
reduction in exchange energy. Interfacial roughness and possibly
interdiffusion of F into Nb is assumed to account for the slightly
smaller exchange energy. Moreover, in particular for Co and Fe,
which exhibit oscillations all in the clean limit, the high exchange
energy means small period of $I_CR_N$ oscillations, enabling us
to obtain the switch from $0$ to $\pi$ state in a very small range
of thicknesses ($\lesssim 3$ nm). We can therefore conclude that Co
and Fe barrier based Josephson junctions are viable structures to
the development of superconductor-based quantum electronic devices.
The electrical and magnetic properties of Co and Fe are well
understood and are routinely used in the magnetics industry: we have
demonstrated that Nb/Co/Nb and Nb/Fe/Nb hybrids can readily be used
in controllable two-level quantum information systems.

\begin{table}[!ht]
\centering
 \caption{\textbf{\label{summary}Summary of the exchange energy $E_{ex}$ and the Fermi velocity $v_F$ extrapolated from theoretical fits of the
  critical current oscillations and the magnetic dead layer $t_D$.} }\bigskip
\begin{tabular}{ c c c c }
  \hline \hline
   Barrier& $E_{ex}(meV)$ & $v_F$($\times 10^5$ m/s) & $t_D$(nm) \\
  \hline
  Ni & 80 & 2.8 & 1.7 \\
  Py & 201 & 2.8 & 0.5\\
  Co & 309 & 2.8 & 0.7\\
  Fe & 256 & 1.98 & 1.1\\
  \hline \hline
\end{tabular}
\vspace*{0.5cm}
\end{table}

 In this context, a feasible future development will be to realize pseudo
spin valve devices using two F layers made of different
ferromagnetic materials with the different coercive fields
separated by a superconducting layer \cite{ChrisPSV}. In this case
the spacer layer is relatively thick, and is used to decouple the
ferromagnetic layers to prevent them from switching at the same
field. In such an instance the fundamental physics will play with
the interlink between these two competing orders, and on the
practical point of view it could be the starting brick for suitable
devices. In this way the magnetization orientation of the F/S/F
structure can be controlled by a weak magnetic field which by itself
is insufficient to destroy superconductivity, but the small magnetic
field enables the softer F layer to be aligned with it, while the
harder material will not switch. This artificial structure allows
active control of the magnetic state of the barrier. The
magnetoresistance of the pseudo spin valve gives direct access to the information
about relative orientation of the ferromagnetic layers, and the
magnetic state of the barrier. The novelty, and strength of such
devices, lies in the fact that these heterostructures can be
realized either with classical ferromagnetic materials and
low-T$_{\rm C}$ superconductors, or with colossal magneto-resistance
materials and high-T$_{\rm C}$ cuprates. So what will be the most
promising future technology??? Will it rely on liquid helium, or
nitrogen???? We hope to contribute towards providing a proper answer
to such dilemmas. On a far-reaching scope, the route to
room-temperature applications would constitute a more formidable
challenge for our research.

\subsection*{Acknowledgements}
I would like to thank M. G. Blamire, J.W.A. Robinson and G. Burnell
for sharing with me the joy of working together on these important
topics presented in this chapter and I thank M. G. Blamire and the
University of Cambridge for the kind hospitality. I acknowledge the
support of the European Science Foundation $\pi$-shift network. Last
but not least, I thank Gerardo Adesso for his support and always
active collaboration and Giovanni Battista Adesso for his brightness
while I wrote this chapter. At the end thanks to Blinky Bill!!


\label{lastpage-01}


\begin{thebibliography}{10}
\expandafter\ifx\csname url\endcsname\relax
  \def\url#1{\texttt{#1}}\fi
\expandafter\ifx\csname urlprefix\endcsname\relax\def\urlprefix{URL }\fi
\providecommand{\bibinfo}[2]{#2}
\providecommand{\eprint}[2][]{\url{#2}}

\bibitem{Samthesis}
\bibinfo{author}{Piano, S.}
\newblock Ph.D. thesis, \bibinfo{school}{University of Salerno,}
  \bibinfo{year}{2007}.

\bibitem{Buzdin}
\bibinfo{author}{Buzdin, A.~I.}
\newblock \emph{\bibinfo{journal}{Rev. Mod. Phys.}}  \bibinfo{year}{2005},
  {\bibinfo{volume}{77}}, \bibinfo{pages}{935}.

\bibitem{Soulen}
\bibinfo{author}{Soulen, R.}; \bibinfo{author}{Byers, J.};
  \bibinfo{author}{Osofsky, M.}; \bibinfo{author}{Nadgorny, B.};
  \bibinfo{author}{T.~Ambrose, S.~C.}; \bibinfo{author}{Broussard, P.};
  \bibinfo{author}{Tanaka, C.}; \bibinfo{author}{Nowak, J.};
  \bibinfo{author}{Moodera, J.}; \bibinfo{author}{Barry, A.};
  \bibinfo{author}{Coey, J.}
\newblock \emph{\bibinfo{journal}{Science}}  \bibinfo{year}{1998},
  {\bibinfo{volume}{282}}, \bibinfo{pages}{85}.

\bibitem{Strijkers}
\bibinfo{author}{Strijkers, G.~J.}; \bibinfo{author}{Ji, Y.};
  \bibinfo{author}{Yang, F.~Y.}; \bibinfo{author}{Chien, C.~L.};
  \bibinfo{author}{Byers, J.~M.}
\newblock \emph{\bibinfo{journal}{Phys. Rev. B}}  \bibinfo{year}{2001},
  {\bibinfo{volume}{63}}, \bibinfo{pages}{104510}.

\bibitem{B2}
\bibinfo{author}{Buzdin, A.~I.}; \bibinfo{author}{Kuprianov, M.~V.}
\newblock \emph{\bibinfo{journal}{JETP Lett.}}  \bibinfo{year}{1990},
  {\bibinfo{volume}{52}}, \bibinfo{pages}{488}.

\bibitem{B3}
\bibinfo{author}{Buzdin, A.~I.}; \bibinfo{author}{Kuprianov, M.~V.}
\newblock \emph{\bibinfo{journal}{JETP Lett.}}  \bibinfo{year}{1991},
  {\bibinfo{volume}{53}}, \bibinfo{pages}{321}.

\bibitem{R1}
\bibinfo{author}{Radovic, Z.};
  \bibinfo{author}{Dobrosavljevi$\acute{c}$-Gruji$\acute{c}$, L.};
  \bibinfo{author}{Buzdin, A.~I.}; \bibinfo{author}{Clem, J.~R.}
\newblock \emph{\bibinfo{journal}{Phys. Rev. B}}  \bibinfo{year}{1988},
  {\bibinfo{volume}{38}}, \bibinfo{pages}{2388}.

\bibitem{R2}
\bibinfo{author}{Radovic, Z.}; \bibinfo{author}{Ledvij, M.};
  \bibinfo{author}{Dobrosavljevi$\acute{c}$-Gruji$\acute{c}$, L.};
  \bibinfo{author}{Buzdin, A.~I.}; \bibinfo{author}{Clem, J.~R.}
\newblock \emph{\bibinfo{journal}{Phys. Rev. B}}  \bibinfo{year}{1991},
  {\bibinfo{volume}{44}}, \bibinfo{pages}{759}.

\bibitem{Demler}
\bibinfo{author}{Demler, E.~A.}; \bibinfo{author}{G.~B.~Arnold, M. R.~B.}
\newblock \emph{\bibinfo{journal}{Phys. Rev. B}}  \bibinfo{year}{1997},
  {\bibinfo{volume}{55}}, \bibinfo{pages}{15 174}.

\bibitem{Andreev}
\bibinfo{author}{Andreev, A.~F.}
\newblock \emph{\bibinfo{journal}{Zh. Eksp. Teor. Fiz.}}  \bibinfo{year}{1964},
  {\bibinfo{volume}{46}}, \bibinfo{pages}{1128}.

\bibitem{Kontos}
\bibinfo{author}{Kontos, T.}; \bibinfo{author}{Aprili, M.};
  \bibinfo{author}{Lesueur, J.}; \bibinfo{author}{Gen\^{e}t, F.};
  \bibinfo{author}{Stephanidis, B.}; \bibinfo{author}{Boursier, R.}
\newblock \emph{\bibinfo{journal}{Phys. Rev. Lett.}}  \bibinfo{year}{2001},
  {\bibinfo{volume}{86}}, \bibinfo{pages}{304}.

\bibitem{Kontos2002}
\bibinfo{author}{Kontos, T.}; \bibinfo{author}{Aprili, M.};
  \bibinfo{author}{Lesueur, J.}; \bibinfo{author}{Gen\^{e}t, F.};
  \bibinfo{author}{Stephanidis, B.}; \bibinfo{author}{Boursier, R.}
\newblock \emph{\bibinfo{journal}{Phys. Rev. Lett.}}  \bibinfo{year}{2002},
  {\bibinfo{volume}{89}}, \bibinfo{pages}{137007}.

\bibitem{Buzdin82}
\bibinfo{author}{Buzdin, A.~I.}; \bibinfo{author}{Bulaevskii, L.};
  \bibinfo{author}{Panyukov, S.}
\newblock \emph{\bibinfo{journal}{JETP Lett.}}  \bibinfo{year}{1982},
  {\bibinfo{volume}{35}}, \bibinfo{pages}{179}.

\bibitem{Bergeret}
\bibinfo{author}{Bergeret, F.~S.}; \bibinfo{author}{Volkov, A.~F.};
  \bibinfo{author}{Efetov, K.~B.}
\newblock \emph{\bibinfo{journal}{Phys. Rev. B}}  \bibinfo{year}{2001},
  {\bibinfo{volume}{64}}, \bibinfo{pages}{134506}.

\bibitem{pijunctionPRB}
\bibinfo{author}{Robinson, J. W.~A.}; \bibinfo{author}{Piano, S.};
  \bibinfo{author}{Burnell, G.}; \bibinfo{author}{Bell, C.};
  \bibinfo{author}{Blamire, M.~G.}
\newblock \emph{\bibinfo{journal}{Phys. Rev. B}}  \bibinfo{year}{2007},
  {\bibinfo{volume}{76}}, \bibinfo{pages}{094522}.

\bibitem{pijunctionPRL}
\bibinfo{author}{Robinson, J. W.~A.}; \bibinfo{author}{Piano, S.};
  \bibinfo{author}{Burnell, G.}; \bibinfo{author}{Bell, C.};
  \bibinfo{author}{Blamire, M.~G.}
\newblock \emph{\bibinfo{journal}{Phys. Rev. Lett.}}  \bibinfo{year}{2006},
  {\bibinfo{volume}{97}}, \bibinfo{pages}{177003}.

\bibitem{IronEPJ}
\bibinfo{author}{Piano, S.}; \bibinfo{author}{Robinson, J. W.~A.};
  \bibinfo{author}{Burnell, G.}; \bibinfo{author}{Bell, C.};
  \bibinfo{author}{Blamire, M.~G.}
\newblock \emph{\bibinfo{journal}{Eur. Phys. J. B}}  \bibinfo{year}{2007},
  {\bibinfo{volume}{58}}, \bibinfo{pages}{123}.

\bibitem{Kim}
\bibinfo{author}{Kim, S.-J.}; \bibinfo{author}{Latyshev, Y.~I.};
  \bibinfo{author}{Yamashita, T.}
\newblock \emph{\bibinfo{journal}{Appl. Phys. Lett.}}  \bibinfo{year}{1999},
  {\bibinfo{volume}{74}}, \bibinfo{pages}{1156}.

\bibitem{Chrisnano}
\bibinfo{author}{Bell, C.}; \bibinfo{author}{Burnell, G.};
  \bibinfo{author}{Kang, D.-J.}; \bibinfo{author}{Hadfield, R.~H.};
  \bibinfo{author}{Kappers, M.~J.}; \bibinfo{author}{Blamire, M.~G.}
\newblock \emph{\bibinfo{journal}{Nanotechnology}}  \bibinfo{year}{2003},
  {\bibinfo{volume}{14}}, \bibinfo{pages}{630}.

\bibitem{Blamire_Eucas}
\bibinfo{author}{Blamire, M.~G.}
\newblock \emph{\bibinfo{journal}{Supercond. Sci. Technol.}}
  \bibinfo{year}{2006}, {\bibinfo{volume}{19}}, \bibinfo{pages}{S132}.

\bibitem{pijunctionIEEE}
\bibinfo{author}{Robinson, J. W.~A.}; \bibinfo{author}{Piano, S.};
  \bibinfo{author}{Burnell, G.}; \bibinfo{author}{Bell, C.};
  \bibinfo{author}{Blamire, M.~G.}
\newblock \emph{\bibinfo{journal}{IEEE Trans. Appl. Supercond.}}
  \bibinfo{year}{2007}, {\bibinfo{volume}{17}}, \bibinfo{pages}{641}.

\bibitem{cbellprb05}
\bibinfo{author}{Bell, C.}; \bibinfo{author}{Loloee, R.};
  \bibinfo{author}{Burnell, G.}; \bibinfo{author}{Blamire, M.~G.}
\newblock \emph{\bibinfo{journal}{Phys. Rev. B}}  \bibinfo{year}{2005},
  {\bibinfo{volume}{71}}, \bibinfo{pages}{180501(R)}.

\bibitem{Born2006}
\bibinfo{author}{Born, F.}; \bibinfo{author}{Siegel, M.};
  \bibinfo{author}{Hollmann, E.~K.}; \bibinfo{author}{Braak, H.};
  \bibinfo{author}{Golubov, A.~A.}; \bibinfo{author}{Gusakova, D.~Y.};
  \bibinfo{author}{Kupriyanov, M.~Y.}
\newblock \emph{\bibinfo{journal}{Phys. Rev. B}}  \bibinfo{year}{2006},
  {\bibinfo{volume}{74}}, \bibinfo{pages}{140501(R)}.

\bibitem{Kitada}
\bibinfo{author}{Kitada, M.}; \bibinfo{author}{Shimizu, N.}
\newblock \emph{\bibinfo{journal}{J. Mat. Sci. Lett.}}  \bibinfo{year}{1991},
  {\bibinfo{volume}{10}}, \bibinfo{pages}{437}.

\bibitem{JAarts}
\bibinfo{author}{Aarts, J.}; \bibinfo{author}{Geers, J. M.~E.};
  \bibinfo{author}{Brück, E.}; \bibinfo{author}{Golubov, A.~A.};
  \bibinfo{author}{Coehoorn, R.}
\newblock \emph{\bibinfo{journal}{Phys. Rev. B}}  \bibinfo{year}{1997},
  {\bibinfo{volume}{56}}, \bibinfo{pages}{2779}.

\bibitem{Renjun}
\bibinfo{author}{Zhang, R.}; \bibinfo{author}{Willis, R.~F.}
\newblock \emph{\bibinfo{journal}{Phys. Rev. Lett.}}  \bibinfo{year}{2001},
  {\bibinfo{volume}{86}}, \bibinfo{pages}{2665}.

\bibitem{Qunwen}
\bibinfo{author}{Leng, Q.}; \bibinfo{author}{Han, H.}; \bibinfo{author}{Hiner,
  C.}
\newblock \emph{\bibinfo{journal}{J. Appl. Phys.}}  \bibinfo{year}{2000},
  {\bibinfo{volume}{87}}, \bibinfo{pages}{6621}.

\bibitem{Pick}
\bibinfo{author}{Pick, S.}; \bibinfo{author}{Turek, I.};
  \bibinfo{author}{Dreyss\'{e}, H.}
\newblock \emph{\bibinfo{journal}{Solid State Commun.}}  \bibinfo{year}{2002},
  {\bibinfo{volume}{124}}, \bibinfo{pages}{21}.

\bibitem{Slater}
\bibinfo{author}{Slater, J.~C.}
\newblock \emph{\bibinfo{journal}{Phys. Rev.}}  \bibinfo{year}{1936},
  {\bibinfo{volume}{49}}, \bibinfo{pages}{537}.

\bibitem{Pauling}
\bibinfo{author}{Pauling, L.}
\newblock \emph{\bibinfo{journal}{J. Appl. Phys.}}  \bibinfo{year}{1937},
  {\bibinfo{volume}{8}}, \bibinfo{pages}{385}.

\bibitem{Blum}
\bibinfo{author}{Blum, Y.}; \bibinfo{author}{Tsukernik, A.};
  \bibinfo{author}{Karpovski, M.}; \bibinfo{author}{Palevski, A.}
\newblock \emph{\bibinfo{journal}{Phys. Rev. Lett.}}  \bibinfo{year}{2002},
  {\bibinfo{volume}{89}}, \bibinfo{pages}{187004}.

\bibitem{Heinmann}
\bibinfo{author}{Heinmann, P.}; \bibinfo{author}{Himpsel, F.};
  \bibinfo{author}{Eastman, D.}
\newblock \emph{\bibinfo{journal}{Solid State Commun.}}  \bibinfo{year}{1981},
  {\bibinfo{volume}{39}}, \bibinfo{pages}{219}.

\bibitem{literature}
\bibinfo{author}{Covo, M.~K.}; \bibinfo{author}{Molvik, A.~W.};
  \bibinfo{author}{Friedman, A.}; \bibinfo{author}{Westenskow, G.};
  \bibinfo{author}{Barnard, J.~J.}; \bibinfo{author}{Cohen, R.};
  \bibinfo{author}{Seidl, P.~A.}; \bibinfo{author}{Kwan, J.~W.};
  \bibinfo{author}{Logan, G.}; \bibinfo{author}{Baca, D.};
  \bibinfo{author}{Bieniosek, F.}; \bibinfo{author}{Celata, C.~M.};
  \bibinfo{author}{Vay, J.-L.}; \bibinfo{author}{Vujic, J.~L.}
\newblock \emph{\bibinfo{journal}{Phys. Rev. ST AB}}  \bibinfo{year}{2006},
  {\bibinfo{volume}{9}}, \bibinfo{pages}{063201}.

\bibitem{Gusakova}
\bibinfo{author}{Gusakova, D.~Y.}; \bibinfo{author}{Kupriyanov, M.~Y.};
  \bibinfo{author}{Golubov, A.~A.}
\newblock \emph{\bibinfo{journal}{JETP Lett.}}  \bibinfo{year}{2006},
  {\bibinfo{volume}{83}}, \bibinfo{pages}{487}.

\bibitem{Gurney}
\bibinfo{author}{Gurney, B.~A.}; \bibinfo{author}{Speriosu, V.~S.};
  \bibinfo{author}{Nozieres, J.~P.}; \bibinfo{author}{Lefakis, H.};
  \bibinfo{author}{Wilhoit, D.~R.}; \bibinfo{author}{Need, O.~U.}
\newblock \emph{\bibinfo{journal}{Phys. Rev. Lett.}}  \bibinfo{year}{1993},
  {\bibinfo{volume}{71}}, \bibinfo{pages}{4023}.

\bibitem{Gil}
\bibinfo{author}{K\"{o}tzler, J.}; \bibinfo{author}{Gil, W.}
\newblock \emph{\bibinfo{journal}{Phys. Rev. B}}  \bibinfo{year}{2005},
  {\bibinfo{volume}{72}}, \bibinfo{pages}{060412(R)}.

\bibitem{Barone}
\bibinfo{author}{Barone, A.}; \bibinfo{author}{Patern\`{o}, G.}
\newblock \bibinfo{note}{\emph{Physics and Applications of the Josephson
  effect}; John Wiley \& Sons: New York, US, 1982}.

\bibitem{Sellier}
\bibinfo{author}{Sellier, H.}; \bibinfo{author}{Baraduc, C.};
  \bibinfo{author}{Lefloch, F.}; \bibinfo{author}{Calemczuk, R.}
\newblock \emph{\bibinfo{journal}{Phys. Rev. Lett.}}  \bibinfo{year}{2004},
  {\bibinfo{volume}{92}}, \bibinfo{pages}{257005}.

\bibitem{ChrisPSV}
\bibinfo{author}{Bell, C.}; \bibinfo{author}{Burnell, G.};
  \bibinfo{author}{Leung, C.~W.}; \bibinfo{author}{Tarte, E.~J.};
  \bibinfo{author}{Kang, D.-J.}; \bibinfo{author}{Blamire, M.~G.}
\newblock \emph{\bibinfo{journal}{Appl. Phys. Lett.}}  \bibinfo{year}{2003},
  {\bibinfo{volume}{84}}, \bibinfo{pages}{1153}.

\end{thebibliography}
\end{document}